\documentclass[%
 reprint,
superscriptaddress,
 amsmath,amssymb,
 aps,
]{revtex4-1}

\usepackage{graphicx}
\usepackage{dcolumn}
\usepackage{bm}
\usepackage{siunitx}
\usepackage{hyperref}

\begin{document}

\title{Nonlocal conductance spectroscopy of Andreev bound states \\ in gate-defined InAs/Al nanowires}

\author{Andreas P\"oschl}
\affiliation{Center for Quantum Devices, Niels Bohr Institute, University of Copenhagen, 2100 Copenhagen, Denmark}

\author{Alisa Danilenko}
\affiliation{Center for Quantum Devices, Niels Bohr Institute, University of Copenhagen, 2100 Copenhagen, Denmark}

\author{Deividas Sabonis}
\affiliation{Center for Quantum Devices, Niels Bohr Institute, University of Copenhagen, 2100 Copenhagen, Denmark}
\affiliation{Laboratory for Solid State Physics, ETH Z\"urich, CH-8093 Z\"urich, Switzerland}%

\author{Kaur Kristjuhan}
\affiliation{Center for Quantum Devices, Niels Bohr Institute, University of Copenhagen, 2100 Copenhagen, Denmark}

\author{Tyler Lindemann}
\affiliation{Department of Physics and Astronomy, and Birck Nanotechnology Center, Purdue University, West Lafayette, Indiana 47907 USA}

\author{Candice Thomas}
\affiliation{Department of Physics and Astronomy, and Birck Nanotechnology Center, Purdue University, West Lafayette, Indiana 47907 USA}

\author{Michael J. Manfra}
\affiliation{Department of Physics and Astronomy, and Birck Nanotechnology Center, Purdue University, West Lafayette, Indiana 47907 USA}
\affiliation{School of Materials Engineering, and School of Electrical and Computer Engineering, Purdue University, West Lafayette, Indiana 47907 USA}

\author{Charles M. Marcus}
\affiliation{Center for Quantum Devices, Niels Bohr Institute, University of Copenhagen, 2100 Copenhagen, Denmark}


\begin{abstract}
The charge character of Andreev bound states (ABSs) in a three-terminal semiconductor-superconductor hybrid nanowire was measured using local and nonlocal tunneling spectroscopy. The device is fabricated using an epitaxial InAs/Al two-dimensional heterostructure with several gate-defined side probes. ABSs are found to oscillate around zero as a function of gate voltage, with modifications of their charge consistent with theoretical expectations for the total Bardeen-Cooper-Schrieffer (BCS) charge of ABSs.

\end{abstract}

\maketitle

In semiconducting nanowires (NWs) proximitized by a layer of superconductor, tunneling spectroscopy from a normal-metal contact reveals a spectrum of particle-hole symmetric Andreev bound states (ABSs) localized within the device, confined either by electrostatic gates, device boundaries, or defects. The process of Andreev reflection enables a measurable current in the presence of states below the superconducting gap. This process of quasiparticle (QP) reflection at the boundary between normal and superconducting phase bears similarity with the reflection of photons from a phase-conjugating mirror \cite{beenakker_microjunctions, beenakker_mirror, beenakker_rmt4topoNIS}. Recently, a novel device geometry has been realized that makes it possible to measure tunneling currents at two  normal leads connected to the same proximitized NW while keeping the parent superconductor grounded \cite{gerbold_nonlocal,denise_nl_gapclosing}.

Nonlocal conductance is measured as a differential current response on one probe in response to a differential voltage applied on another probe. For applied voltages smaller than the superconducting gap, nonlocal transport is mediated by Andreev states that couple to the relevant tunnel probes. Theoretical studies predict a characteristic signature in nonlocal conductance of a topological phase transition in NWs with particular spin-orbit and Zeeman effect \cite{andreev_rectifier, karsten_nl_spectroscopy, SDS_nl_conductance, hess_nl_quasimajo}. Characteristic symmetry relations relating local and nonlocal conductances have been reported experimentally \cite{gerbold_nonlocal}. The closing of the induced gap measured in nonlocal conductance in applied field has also been reported experimentally \cite{denise_nl_gapclosing}. Using the same transport processes, quantum dots coupled to one superconducting and two normal leads have been used to demonstrate Cooper-pair splitting \cite{CPS_1, CPS_2, CPS_3, CPS_4}. Nonlocal spectroscopy of subgap states induced by quantum dot states have been reported in vapor-liquid-solid grown NWs and carbon nanotubes \cite{gramich_nl_abs, CPS_3}.

\begin{figure}[b]
\includegraphics[scale=0.9]{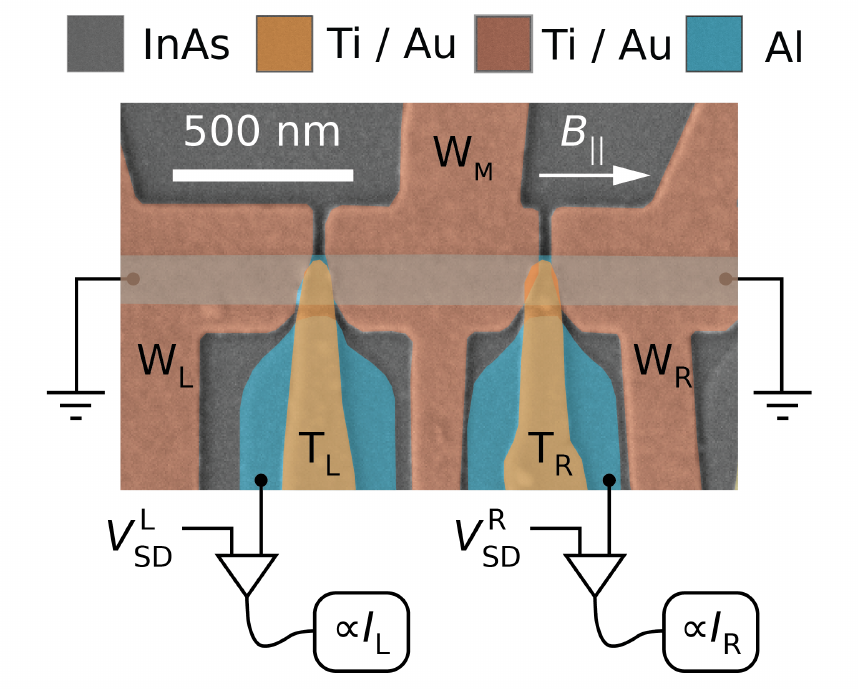}
\caption{\label{fig:device}False-color electron micrograph of device 1. A proximitized quasi-one-dimensional NW is formed in the InAs quantum well (gray) under the strip of superconducting Al (blue) by lateral electrostatic confinement from the gates $\mathrm{W_L,W_M,W_R}$ (red). Probes made from Al are separated by a tunnel barrier from the NW and allow for measurements of the tunneling currents $I_\mathrm{L}$, $I_\mathrm{R}$ into the NW. The gates $\mathrm{T_L}$ and $\mathrm{T_R}$ (orange) tune the tunnel barriers between the NW and the probes.}
\end{figure} 

Here, we investigate and compare local and nonlocal conductance measurements on a gate-defined NW  formed from an InAs two-dimensional electron gas (2DEG) with epitaxial Al. By using a patterned 2DEG, lithographically defined gate-controlled tunneling probes can be coupled laterally to the NW rather than at the ends, allowing several equivalent probes along the side of the NW. We observed characteristic local and nonlocal signatures of ABSs that have been intentionally tuned to couple neighboring probes. Three similar multi-probe devices were measured. Device 1 had superconducting leads that were driven normal by a magnetic field applied along the NW; device 2 had the epitaxial superconductor in the leads removed. Except for supercurrent signatures in local (but not nonlocal) conductance at low fields (below \SI{0.2}{\tesla}) due to superconductivity in the leads, devices 1 and 2 showed similar behavior. A micrograph and data from device 2 are shown in the Supplementary Material (SM). A third measured device was operational but did not show clear ABSs between adjacent probes, and was not investigated further.

A micrograph of device 1 is shown in Fig.~\ref{fig:device}. The device consists of a superconducting strip of Al, defined by wet etching of the epitaxial Al, on top of a shallow 2DEG formed in an InAs quantum well. The strip serves both to induce superconductivity in the InAs by proximity, and to define the wire width by screening the surrounding gates. Two superconducting probes patterned in the same lithographic step were defined $\SI{50}{\nano\meter}$ away from the superconducting strip, leaving a narrow gap of exposed semiconductor between the NW and the probe. Ti/Au gates insulated by $\mathrm{HfO_x}$ gate dielectric were used to control the probe conductance and the electrostatic environment around the NW. Gates labeled $\mathrm{W_L}$, $\mathrm{W_M}$, and $\mathrm{W_R}$ cover different segments of the Al strip and electrostatically confine an approximately $\SI{100}{\nano\meter}$ wide NW. Gates $\mathrm{T_L}$ and $\mathrm{T_R}$ deplete the semiconductor between NW and the respective left and right probe, controlling the tunnel barrier. The ends of the superconducting strip were connected to ground planes of superconducting Al.

Separate current-to-voltage converters on two tunnel probes allow simultaneous measurement of currents $I_\mathrm{L}$ and $I_\mathrm{R}$ as a function of source-drain bias voltages $V^\mathrm{L}_\mathrm{SD}$ and $V_\mathrm{SD}^\mathrm{R}$ (positive current is defined as flowing from the amplifier to the device). Lock-in detection following Ref.~\cite{gerbold_nonlocal} was used to measure the local and nonlocal differential tunneling conductances
\begin{align}\label{eq:Gmat}
    G_\mathrm{LL}=\mathrm{d}I_\mathrm{L}/\mathrm{d}V^\mathrm{L}_\mathrm{SD} && G_\mathrm{LR} = \mathrm{d}I_\mathrm{L}/\mathrm{d}V^\mathrm{R}_\mathrm{SD}
\end{align}
and $G_\mathrm{RR}$, $G_\mathrm{RL}$ defined analogously. 
Further details are given in the SM. \\

\begin{figure}[t]
\includegraphics[scale=0.9]{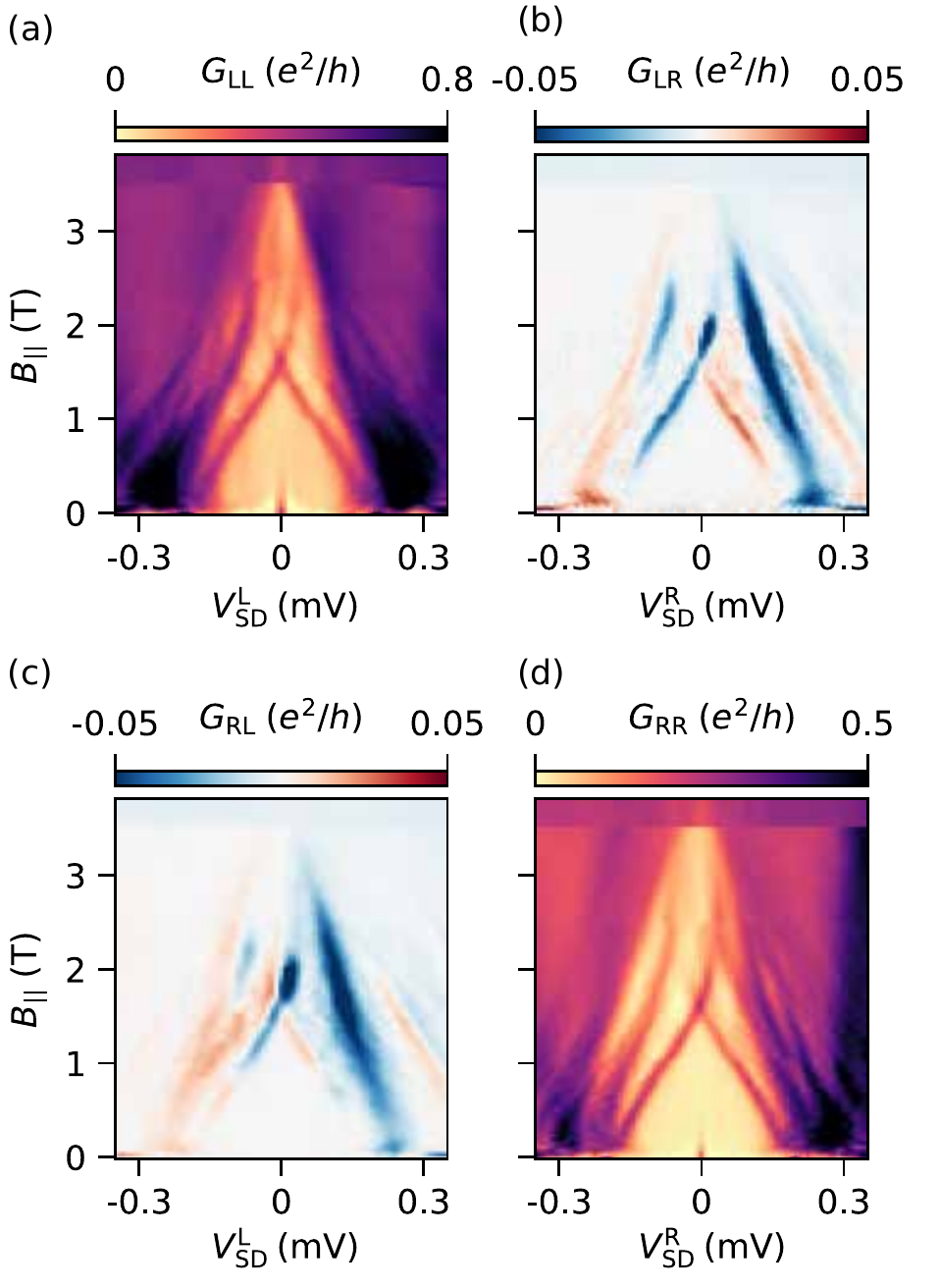}
\caption{\label{fig:fieldscan}Local (LL and RR) and nonlocal (LR and RL) differential conductances as a function of magnetic field $B_{||}$ parallel to the NW, at $V_\mathrm{WM}=\SI{-3.02}{\volt}$ (see extended ranges of $V_\mathrm{WM}$ in Fig.~\ref{fig:plungerscans}). (a, d) Local conductances reveal subgap states that cross zero bias. (b, c) Subgap states also appear in the nonlocal conductances. Note that local differential conductances are positive everywhere, while nonlocal conductances are roughly balanced around zero.
}
\end{figure}

To confine an ABS in a segment of NW, a modulation of the potential along the NW was created using gates $\mathrm{W_L,\,W_M,\,W_R}$. Setting $V_\mathrm{WL}$ and $V_\mathrm{WR}$ to \SI{-4.50}{\volt} created a hard, superconducting gap with no subgap states in these segments. The middle gate voltage $V_\mathrm{WM}$ was then set to $\SI{-3.02}{\volt}$, less negative than its neighboring gates. Local and nonlocal conductances (\ref{eq:Gmat}) as a function of magnetic field along the NW, $B_{||}$, is shown in Fig.~\ref{fig:fieldscan}.  Above $B_{||}>\SI{0.2}{\tesla}$ the Al leads are driven normal, providing normal-metal probes \cite{henri_lead}. Both local tunneling conductances $G_\mathrm{LL}$ and $G_\mathrm{RR}$ show subgap resonances that emerge at low magnetic fields from the continuum at high bias and cross zero voltage bias at $B_{||}=\SI{1.6}{\tesla}$. We associate these resonances with an extended ABS in the \SI{0.6}{\micro\meter} long NW segment under gate $\mathrm{W_M}$ due to the appearance in both local tunneling conductances with identical dependence on magnetic field and gate voltage $V_\mathrm{WM}$, discussed below. Hybridization of the ABS with an accidental resonance in one of the tunnel barriers was reported previously \cite{PoschlPC2022}.

Nonlocal conductances $G_\mathrm{LR}$ and $G_\mathrm{RL}$ were of order $ 5\times10^{-2} \,e^2/h$, roughly a factor of 10 smaller than corresponding local conductances. Comparing these values to simulation suggests low-to-moderate disorder \cite{SDS_nl_conductance}. Larger nonlocal conductance, $G_\mathrm{LR},G_\mathrm{RL} > 2 \times 10^{-2} \, e^2/h$ at higher bias presumably reflect the field-dependent superconducting gap, $\Delta(B_{||})$, of the Al layer. For magnetic fields below the zero crossing of the ABSs, there is a region of unmeasurably small nonlocal conductance around zero bias, which extends to $V^\mathrm{L/R}_\mathrm{SD}$ values marking the ABS energy. The ABSs are the lowest lying excited states that extend over the full segment of NW under gate $\mathrm{W_M}$. Their energy therefore sets the size of the energy gap  $\Delta_{\mathrm{ind}}$ that is induced in the semiconductor by proximity effect and Zeeman energy. Exponentially suppressed nonlocal conductance is expected for $eV^\mathrm{L/R}_\mathrm{SD}<\Delta_\mathrm{ind}$ for NWs that are longer than the wavefunction decay length at these energies \cite{andreev_rectifier}. In the voltage range $\Delta_\mathrm{ind}< eV_\mathrm{SD}\leq\Delta$ nonlocal conductance can be interpreted as QP transport through excited states.

\begin{figure}
\includegraphics[scale=1.0]{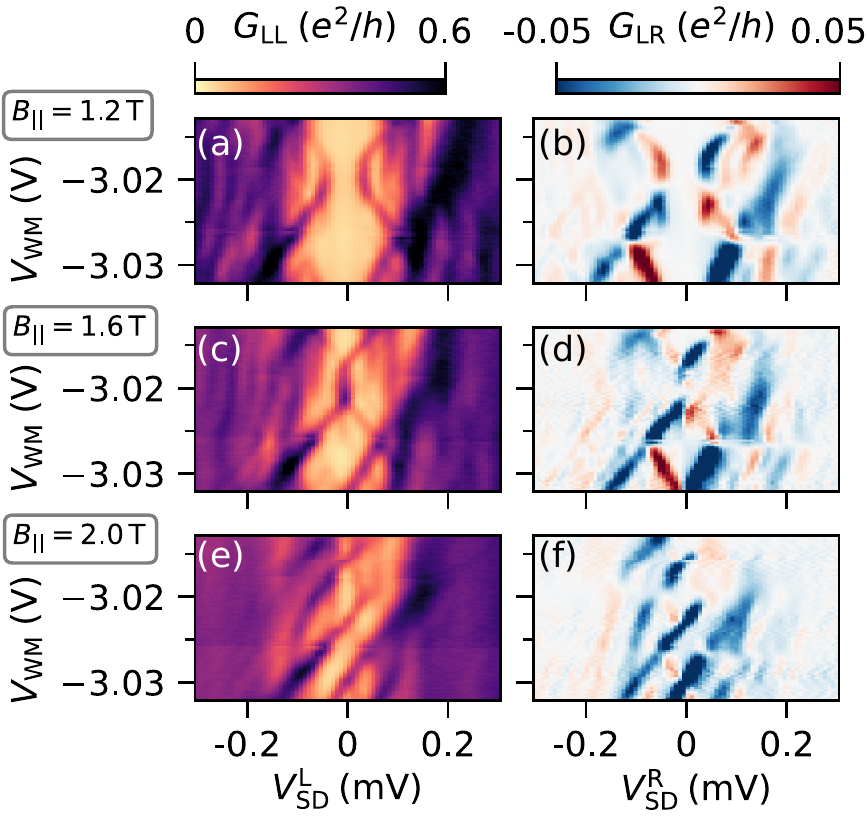}
\caption{\label{fig:plungerscans}Local conductance $G_\mathrm{LL}$ and nonlocal conductance $G_\mathrm{LR}$ as a function of $V_\mathrm{WM}$ at three different values of magnetic field $B_{||}$. (a) Subgap states appear as lobes in the superconducting gap. (b) The lowest excited state appears together with a spectrum of higher excited states in the nonlocal conductance. (c) At $B_{||}=\SI{1.6}{\tesla}$ the ABSs merge at zero bias. (d) The nonlocal conductance is suppressed around that region. (e) At $B_{||}=\SI{2}{\tesla}$ the ABSs intersect forming a low energy state that oscillates around zero bias. (f) The nonlocal conductance changes sign at the turning points of the low energy state. }
\end{figure}

At a magnetic field of $B_{||}=\SI{1.2}{\tesla}$, the ABSs trace out a pair of lobes that do not reach zero bias as a function of gate voltage $V_\mathrm{WM}$, as seen in the the local conductance $G_\mathrm{LL}$ [see Fig.~\ref{fig:plungerscans}(a)]. The corresponding nonlocal conductance  $G_\mathrm{LR}$, plotted in Fig.~\ref{fig:plungerscans}(b), is largest at a value $V_\mathrm{SD}^\mathrm{R}$ that tracks the position of the low energy subgap state in $G_\mathrm{LL}$. Following this state, the nonlocal conductance changes sign in two cases. The first case is a value $V_\mathrm{WM}$ at which the ABS reaches a minimum in energy. The second case are points where two ABSs cross, which leads to the energy of the lowest lying state changing its slope abruptly from positive to negative and vice versa. Note that there is a spectrum of additional excited states visible at higher bias values $V_\mathrm{SD}^\mathrm{R}$. We interpret these states as a result of the NW being sufficiently long such that the spacing between excited states is decreased \cite{Mishmash2016}. A similarly dense spectrum of excited states has been absent in nonlocal conductance measurements on proximitized quantum dots \cite{gramich_nl_abs} and shorter NWs \cite{gerbold_nonlocal}. 

At a magnetic field of $B_{||}=\SI{1.6}{\tesla}$ the lowest ABSs merge at zero voltage bias for a small interval of $V_\mathrm{WM}$, as seen in local conductance in Fig.~\ref{fig:plungerscans}(c). Within this range, the nonlocal conductance through the ABS is much smaller than for values of  $V_\mathrm{WM}$ where the ABSs are away from zero bias [see Fig.~\ref{fig:plungerscans}(d)]. This can be understood as a result of the rates for crossed Andreev reflection and QP transmission being equal at this point due to particle-hole symmetry \cite{Akhmerov2011, lobos_topo_pt_NSN}. Nonlocal conductance vanishes at the crossing point as it is proportional to the difference of these two rates \cite{Takane1992,datta_super_LBK,lobos_topo_pt_NSN}. At a magnetic field $B_{||}=\SI{2}{\tesla}$ the ABSs intersect, creating a low energy state that oscillates around zero bias, as seen in Fig.~\ref{fig:plungerscans}(e). $G_\mathrm{LL}$ shows an asymmetry with respect to $V_\mathrm{SD}^\mathrm{L}$. 

Local tunneling conductances are expected to be symmetric with respect to source-drain bias for the case of two-terminal devices, but for the case of three terminal devices a finite asymmetry is expected. In a linear transport theory, the antisymmetric part of the nonlocal conductances $G_\mathrm{LR}^\mathrm{anti}(V_\mathrm{SD}^\mathrm{R})=[G_\mathrm{LR}(V_\mathrm{SD}^\mathrm{R})-G_\mathrm{LR}(-V_\mathrm{SD}^\mathrm{R})]/2$ fulfills the relations 
\begin{equation}
    G_\mathrm{LR}^\mathrm{anti}(V_\mathrm{SD}^\mathrm{R}) = -G_\mathrm{LL}^\mathrm{anti}(V_\mathrm{SD}^\mathrm{L})
\end{equation}
 at subgap voltages $eV_\mathrm{SD}<\Delta$ as a consequence of particle-hole symmetry and current conservation \cite{karsten_nl_spectroscopy, gerbold_nonlocal}. Analogously, one expects $G_\mathrm{RL}^\mathrm{anti}(V_\mathrm{SD}^\mathrm{L}) = -G_\mathrm{RR}^\mathrm{anti}(V_\mathrm{SD}^\mathrm{R})$. We find that these relations are quantitatively fulfilled for the lowest excited state, while they are violated for higher excited states. Consequently, the sum over all local and nonlocal conductances $G_\mathrm{sum}= G_\mathrm{LL}+G_\mathrm{RR}+G_\mathrm{LR}+G_\mathrm{RL}$ is symmetric up to source-drain bias voltages of the lowest energy state. A detailed analysis can be found in the SM. Possible reasons for deviations from the symmetry relations were given in \cite{gerbold_nonlocal}. In addition, numerical studies have shown that an energy dependence of the tunnel barriers in a nonlinear transport theory can give rise to violations of the symmetry relations \cite{melo_asym}.\\

\begin{figure}
\includegraphics[scale=0.8]{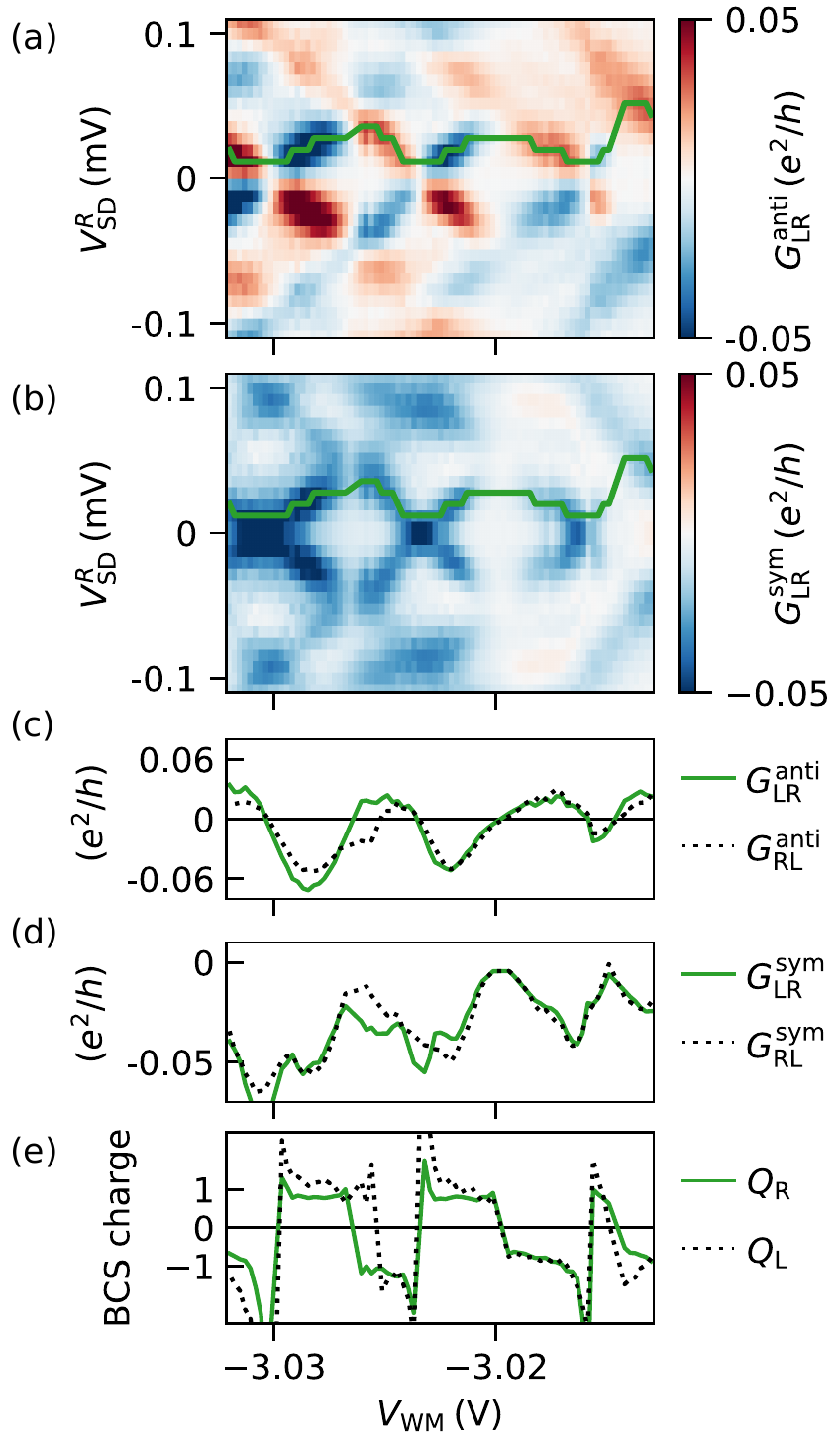}
\caption{\label{fig:Qextraction}(a, b) antisymmetric and symmetric component of the nonlocal conductance $G_\mathrm{LR}$ measured at $B_{||}=\SI{2}{\tesla}$. 
(c, d) antisymmetric and symmetric part of the nonlocal conductances $G_\mathrm{LR}$,\;$G_\mathrm{RL}$ extracted at the position of the lowest lying state marked by the green lines in (a, b). (e) Resulting values for $Q_\mathrm{L}$ and $Q_\mathrm{R}$ are approximately equal and show clear oscillations between +1 and -1. Positive (negative) values of $Q_\mathrm{j}$ coincide with regions of positive (negative) slope of the state energy as a function of $V_\mathrm{WM}$. }
\end{figure}

The quantity
\begin{equation}
 \begin{aligned}\label{eq:BCS_relation}
 Q_\mathrm{R}=\mathrm{sign}(V_\mathrm{SD}^\mathrm{R})\left.\frac{G^\mathrm{sym}_\mathrm{LR}(V_\mathrm{SD}^\mathrm{R})}{G^\mathrm{anti}_\mathrm{LR}(V_\mathrm{SD}^\mathrm{R})}\right\vert_{E=eV_\mathrm{SD}^\mathrm{R}}
 \end{aligned}
\end{equation} of a subgap state at energy $E=eV^\mathrm{R}_\mathrm{SD}$ can be extracted from the antisymmetric and symmetric components of the measured nonlocal conductance $G_\mathrm{LR}(V^\mathrm{R}_\mathrm{SD})$ \cite{gerbold_nonlocal}. An equivalent quantity $Q_\mathrm{L}$ can be defined based on $G_\mathrm{RL}(V^\mathrm{L}_\mathrm{SD})$. The symmetric and antisymmetric components of the nonlocal conductance $G_\mathrm{LR}$ measured at a magnetic field value $B_{||}=\SI{2}{\tesla}$ are plotted as a function of source-drain voltage in Figs.~\ref{fig:Qextraction}(a, b). The values stemming from the low energy state were extracted along the positions given by the green lines in Figs.~\ref{fig:Qextraction}(c, d).

Extracted antisymmetric and symmetric parts of $G_\mathrm{LR}$ (solid green) and $G_\mathrm{RL}$ (dotted black) are shown in Figs.~\ref{fig:Qextraction}(c, d). These values correspond to the conductances that enter the expressions for $Q_\mathrm{R}$ and $Q_\mathrm{L}$. Note that symmetric and antisymmetric parts of $G_\mathrm{LR}$ and $G_\mathrm{RL}$ are roughly equal. The resulting values for $Q_\mathrm{L}$ and $Q_\mathrm{R}$ according to Eq.~(\ref{eq:BCS_relation}) are shown in Fig.~\ref{fig:Qextraction}(e). $Q_\mathrm{L}$ closely follows $Q_\mathrm{R}$.
\\

Theoretically, the values $Q_\mathrm{L}$ and $Q_\mathrm{R}$ are proportional to the local BCS charge of the bound state at the left and right probe position respectively \cite{Hellenes2019, karsten_nl_spectroscopy, gerbold_nonlocal}.
We find that local charge character on the left and the right are approximately equal, $Q_\mathrm{L}\approx Q_\mathrm{R}$. For device 1 and $B_{||}=\SI{2}{\tesla}$, there are extended plateaus $Q_\mathrm{j}\approx+1$ or $Q_\mathrm{j}\approx-1$ ($j\in\{\mathrm{L},\, \mathrm{R}\}$) indicating a state which is locally fully electron or fully hole-like.
Regions of constant positive $Q_\mathrm{j}$ coincide with ranges in $V_\mathrm{WM}$ where the state energy has a positive slope with respect to $V_\mathrm{WM}$. Regions of negative $Q_\mathrm{j}$ appear where the state has a negative slope. Abrupt changes in $Q_\mathrm{j}$ appear at crossing points of states at finite and zero source-drain bias. This is in agreement with the interpretation of $Q_\mathrm{j}$ measuring the local charge of the bound state. For lower magnetic field values, at which the ABSs appear as parabolic lobes without zero energy crossings, a continuous change of  $Q_\mathrm{j}$ from -1 to 1 is found at the point of minimal ABS energy [see Fig.~S6 in SM]. For device 2, the same behavior, namely $Q_\mathrm{L}\approx Q_\mathrm{R}$, was observed, with either abrupt changes or continuous crossover from positive to negative $Q_\mathrm{j}$.

The total, integrated charge of a bound state at energy $E$ is expected to be proportional to $\mathrm{d}E/\mathrm{d}V_\mathrm{WM}$ according to a model based on a Bogoliubov-de Gennes Hamiltonian \cite{karsten_nl_spectroscopy}. Integrating the total charge over a range of gate voltages should therefore recover the energy of the subgap state as a function of $V_\mathrm{WM}$. We numerically integrated the experimentally determined $Q_\mathrm{j}$ after re-scaling by a lever arm $a$ and taking into account a linear background $b$ and integration constant $c$, yielding the inferred energy  
\begin{equation}
    \begin{aligned}
    \widetilde{E}_\mathrm{j}=a\int Q_{j} \mathrm{d}V_\mathrm{WM} + b V_\mathrm{WM} + c.
    \end{aligned}
\end{equation}
$a$, $b$, and $c$ are free parameters. We find that the resulting curves for $\widetilde{E}_\mathrm{L} (V_\mathrm{WM})$ and $\widetilde{E}_\mathrm{R} (V_\mathrm{WM})$ match the energy evolution $E(V_\mathrm{WM})$ of the low energy subgap state over an extended range of $V_\mathrm{WM}$ (see Fig.~S7 in SM). This suggests that the experimentally determined $Q_\mathrm{j}$ not only reflects the local charge character of the ABS but serves as measure for the total charge of the bound states. Deviations from this behavior are expected for longer devices were the QP charge can vary along the spatial extent of bound states \cite{karsten_nl_spectroscopy,Hellenes2019}.

In summary, we have performed local and nonlocal conductance spectroscopy in a 2DEG based nanowire with integrated side probes and multiple confining gates along the nanowire length, as a  function of magnetic field and gate voltage on a middle segment of the nanowire surrounded by hard-gap regions created by more-negative gate voltages in adjacent regions. The predicted symmetry relations between the antisymmetric components of local and nonlocal conductances are fulfilled for the lowest excited state. In addition, we find a dense spectrum of excited states that give rise to nonlocal conductance. For the lowest excited state, the extracted charge character is the same at both NW ends. This is similar to previous studies \cite{gerbold_nonlocal} despite a longer NW being used here. At high magnetic fields the charge character $Q_\mathrm{L}$, $Q_\mathrm{R}$ of the low energy state alternates between fully electron and hole-like. The oscillations in the charge character are found to be in agreement with the energy evolution $E(V_\mathrm{\mathrm{WM}})$ of the subgap state which suggests that $Q_\mathrm{L}$ and $Q_\mathrm{R}$ reflect the total charge of the ABS measured. 

In the SM we show additional data measured on device~1, together with data from a second device (device 2). The SM also contains further details on the materials used, the device fabrication, and the data analysis.

\begin{acknowledgments}
We thank Karsten Flensberg, Max Geier, Torsten Karzig, Andrea Maiani, Dmitry Pikulin, Waldemar Svejstrup, and Georg Winkler for valuable discussions concerning theory, and Abhishek Banerjee, Lucas Casparis, Asbj\o rn Drachmann, Esteban Martinez, Felix Passmann, Daniel Sanchez, Saulius Vaitiek\.enas, and Alexander Whiticar for experimental input. We acknowledge supported from the Danish National Research Foundation, Microsoft, and a grant (project 43951) from VILLUM FONDEN.
\end{acknowledgments}

\bibliography{references_nl}

\appendix
\clearpage
\onecolumngrid

\begin{center}
{\bf SUPPLEMENTARY MATERIAL}
\end{center}

\setcounter{equation}{0}
\setcounter{figure}{0}
\setcounter{table}{0}
\setcounter{page}{1}
\makeatletter
\renewcommand{\theequation}{S\arabic{equation}}
\renewcommand{\thefigure}{S\arabic{figure}}

\section{Materials and device fabrication}

The material for device 1 was an $\mathrm{In_{1-x}Ga_{x}As-InAs-In_{1-x}Ga_{x}As}$ heterostructure with epitaxial Al grown \textit{in situ}. The InAs quantum well was \SI{7}{\nano\meter} thick and the Al was \SI{5}{\nano\meter} thick. The structure was grown on an InP substrate using molecular beam epitaxy. A buffer between the quantum well and the InP was grown to filter dislocation defects and improve the lattice matching. The material for device 2 was based on an $\mathrm{In_{1-y}Al_{y}As-InAs-In_{1-x}Ga_{x}As}$ quantum well (InAs thickness \SI{7}{\nano\meter}) with a 5~nm thick layer of Al. For both wafers the Al was passivated with a layer of native oxide formed in a controlled oxidation succeeding the growth.

The wafer was scribed and cleaved. Device fabrication followed starting with a mesa etch defined by electron beam lithography (EBL). The mesa etch defines the leads, bondpads, and active region on a micrometer scale. It consisted of etching the Al layer, followed by a \SI{350}{\nano\meter} etch using $\mathrm{H_2O:C_6H_8O7:H_3PO_4:H_2O_2}$ (220:55:3:3). The Al film was further removed selectively using EBL and Transene aluminum etchant type D for \SI{5}{\second} at $50^\circ\mathrm{C}$. Gate dielectric $\mathrm{HfO_{x}}$ of thickness $\SI{15}{\nano\meter}$ was deposited globally at $90^\circ\mathrm{C}$ using atomic layer deposition. A first layer of gate electrodes was fabricated in two steps, each consisting of an EBL defined liftoff process. Inner parts of the gate electrodes which cover the active region of the device were fabricated in the first step from \SI{5}{\nano\meter}/\SI{20}{\nano\meter} Ti/Au. In the second step, the outer parts of the gate electrodes and the bond pads were fabricated from \SI{15}{\nano\meter}/\SI{350}{\nano\meter} Ti/Au to ensure that the gates are continuous at the mesa sidewalls. For device 1 a second layer of \SI{12}{\nano\meter} gate dielectric $\mathrm{HfO_x}$ was grown by atomic layer deposition followed by the deposition of a second layer of gate electrodes in a separate lithography step that included only the gates labeled $\mathrm{W}_\mathrm{j}$ (shown in red in Fig.~\ref{fig:detection}). 

\section{Electrical measurement details}
\begin{figure}[h]
\includegraphics[scale=0.6]{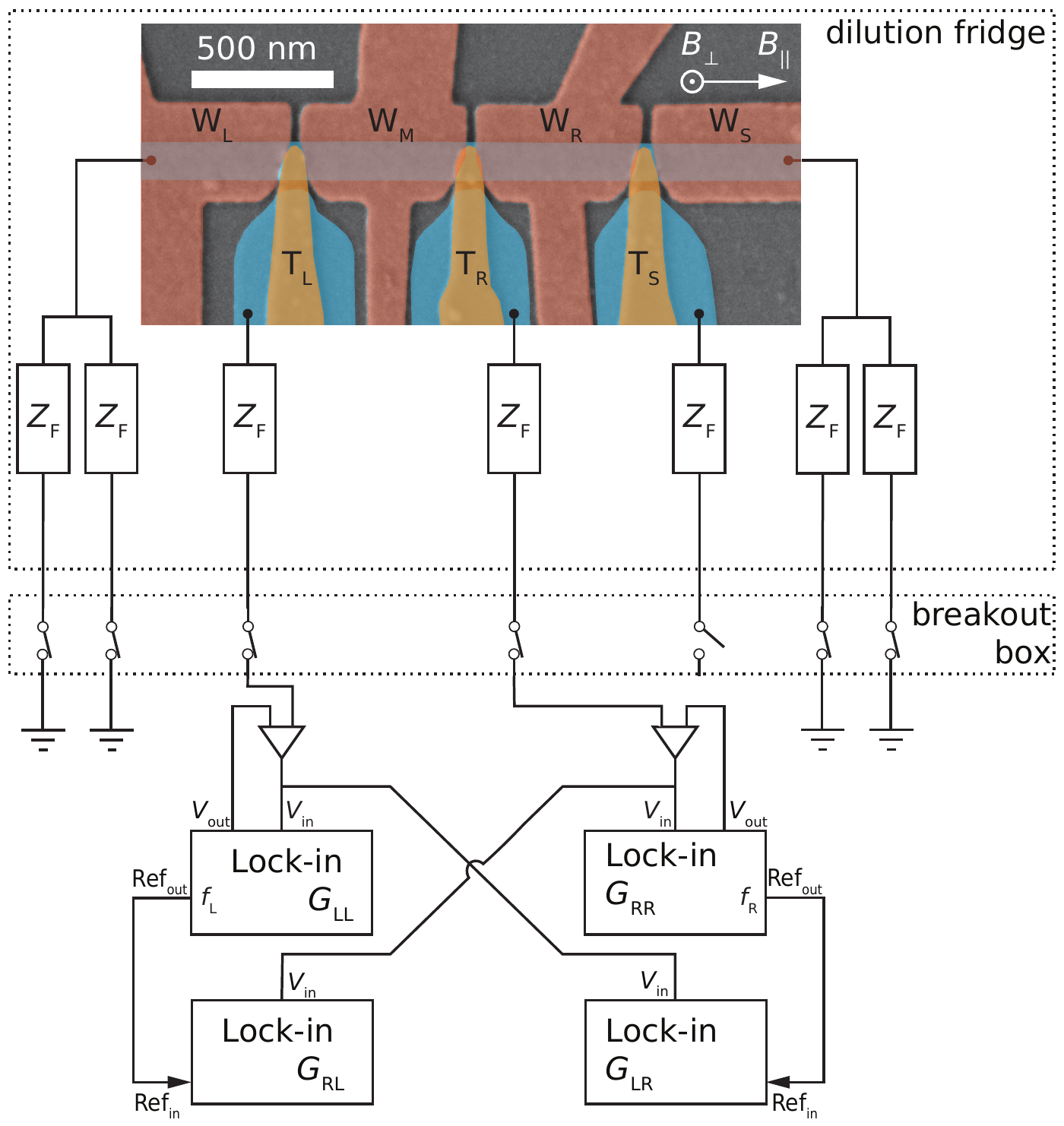}
\caption{\label{fig:detection}Schematic of the electrical measurement setup. The sample was cooled in a dilution refrigerator. Connections to measurement electronics were made by electrical lines of impedance $Z_{\mathrm{F}}$. Two current to voltage amplifiers were used to measure the tunneling current across the tunnel barriers under the gates $\mathrm{T_L}$ and $\mathrm{T_R}$. Four lock-in amplifiers in the shown configuration were used to measure the four elements of the conductance matrix. The lead under the gate $\mathrm{T_S}$ was terminated by an open circuit at the breakout box.}
\end{figure}

The devices were wire-bonded with Al wire and cooled in a cryo-free dilution refrigerator (Oxford Instruments, Triton 400) equipped with a 6-1-\SI{1}{\tesla} vector magnet. The mixing chamber temperature was $\approx\SI{15}{\milli\kelvin}$ as measured by a $\mathrm{RuO_2}$ thermometer.\\
A schematic of the measurement setup is shown in Fig.~\ref{fig:detection}. All electrical lines in the cryostat were equipped with in-house built, multi-stage cryogenic RF and RC filters (cutoff frequency \SI{80}{\mega\hertz} and \SI{0.7}{\kilo\hertz}). The resulting total line impedance $Z_{\mathrm{F}}$ at lock-in frequencies ($<\SI{100}{\hertz}$) is dominated by its resistance \SI{0.88}{\kilo\ohm} stemming from \SI{0.180}{\kilo\ohm} line resistance and \SI{0.70}{\kilo\ohm} filter resistance.  In order to mitigate circuit-effects that can occur for three terminal devices \cite{voltage_divider} a low resistance to ground of the parent superconductor is desirable. This was achieved by bonding each of the ground planes at the respective nanowire (NW) end to two electrical lines. These four lines were connected to ground at the breakout box at room temperature, creating a resistance to ground that is four times smaller compared to a single line. \\

For nonlocal conductance measurements as described in the main text, the detection scheme as shown in the bottom half of Fig.~\ref{fig:detection} was used. The two leads under the gates $\mathrm{T_L}$ and  $\mathrm{T_R}$ were connected to individual low-noise high-stability current to voltage converting amplifiers (Basel Precision Instruments, SP983c) with a gain of $10^8\: \SI{}{\volt}$/A. In addition to the two tunnel probes under the gates $\mathrm{T_L}$ and $\mathrm{T_R}$, the device is equipped with a third lead under the gate $\mathrm{T_S}$ which was terminated with an open circuit at the breakout box for the measurements presented in the main text. Voltage offsets $V_{\mathrm{SD}}^\mathrm{L}$, $V_{\mathrm{SD}}^\mathrm{R}$ at the current input of the amplifiers were applied by using the offset voltage input of the amplifier. The DC component of $V_{\mathrm{SD}}^\mathrm{L}$, $V_{\mathrm{SD}}^\mathrm{R}$ were supplied by an in-house built digital to analog converter while additional AC modulations $\mathrm{d}V_{\mathrm{SD}}^\mathrm{L}$, $\mathrm{d}V_{\mathrm{SD}}^\mathrm{R}$ at reference frequencies $f_\mathrm{L}=\SI{71}{\hertz}$,  $f_\mathrm{R}=\SI{30}{\hertz}$ were supplied by the outputs of lock-in amplifiers (Stanford Research Systems, SR830). The nonlocal conductance was detected by measuring the output of each amplifier with two lock-in amplifiers. One of the lock-in amplifiers was locked to  $f_\mathrm{L}$ while the other was locked to  $f_\mathrm{R}$. The same detection technique has been applied in previous works \cite{denise_nl_gapclosing, gerbold_nonlocal}.

\section{Data analysis}
\subsection{Symmetrization and anti-symmetrization of data}
In order to investigate the relation between the antisymmetric parts of the different conductances with respect to $V_{\mathrm{SD}}$, the antisymmetric parts were extracted from the measured data. The symmetric components with respect to $V_{\mathrm{SD}}$ were further extracted to determine $Q_\mathrm{L}$ and $Q_\mathrm{R}$. \\

Symmetrizing and antisymmetrizing the data is sensitive to small offsets of $V_{\mathrm{SD}}$ from zero, which can occur in experiments despite careful calibration prior to measurements. In order to compensate for voltage offsets, that are smaller than the spacing of measured points in $V_{\mathrm{SD}}$, the data was up-sampled along the $V_{\mathrm{SD}}$ dimension to roughly three times the resolution. The up-sampled data was then symmetrized and antisymmetrized around the symmetry point which is typically offset from zero by two to four pixels in the new, up-sampled dimension. After forming the symmetric and antisymmetric parts of the data, the data was down-sampled again to restore the resolution of the underlying raw data.

\subsection{Extraction of values at peak positions}

For the extraction of the quantity $Q_\mathrm{L}$ of a state at energy $E$, it is necessary to extract the values of $G^\mathrm{sym}_\mathrm{RL}$ and $G^\mathrm{anti}_\mathrm{RL}$ at the voltage $e V_\mathrm{SD}^\mathrm{L}=E$. Typically there is a local extremum in at least one of the measured conductance matrix elements around this voltage value. The routine used to determine this position was to search for local maxima in $|G^\mathrm{anti}_\mathrm{RL}(V_\mathrm{SD}^\mathrm{L})|$, $|G^\mathrm{sym}_\mathrm{RL}(V_\mathrm{SD}^\mathrm{L})|$, or $|G_\mathrm{sum}|=|G_\mathrm{LL}+G_\mathrm{RR}+G_\mathrm{LR}+G_\mathrm{RL}|$ 
for every value of gate voltage. Only one of the three quantities was used depending on which one yielded the best initial guess for the evolution of the state evaluated by eye. 
Points where the automated routine was not accurate (clear by eye) due to the absence of an obvious local maximum were removed by hand. The number of such erroneously detected peak locations was below 5 out of 75 in every case reported here. The analogous procedure is applied for detecting peak positions for the calculation of $Q_\mathrm{R}$.

\section{Supplementary data on Device 1}
\begin{figure}[h]
\includegraphics[scale=0.9]{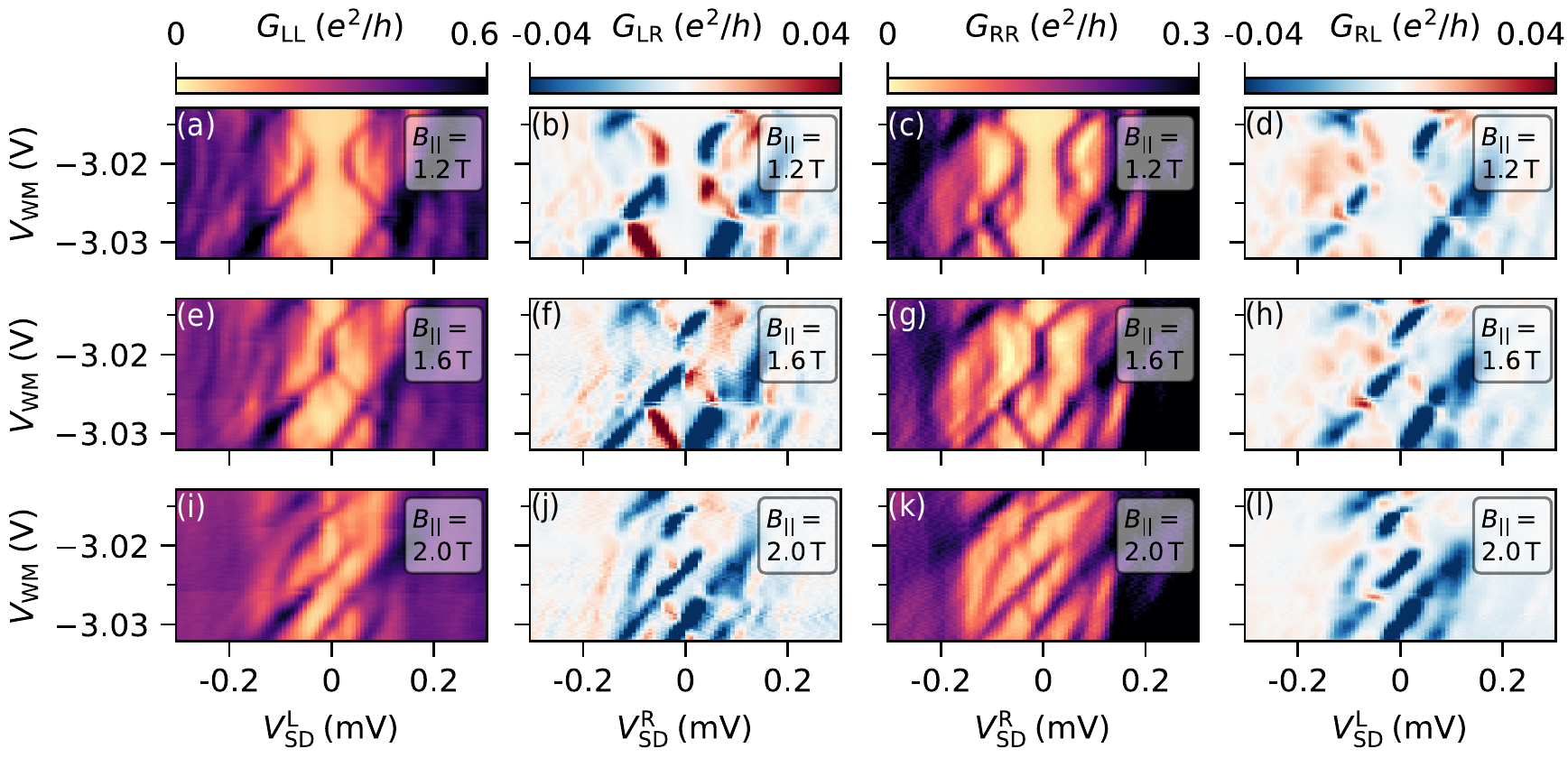}
\caption{\label{fig:full_plunger_dev_1}Each row shows the local and nonlocal conductances as a function of gate voltage $V_\mathrm{WM}$ at $B_{||} = $~1.2~T, 1.6~T, and 2.0~T. The two leftmost columns are the data presented in the main text.}
\end{figure}

The data in the following supplement the main text. In Fig.~\ref{fig:full_plunger_dev_1} we show all four conductances as a function of gate voltage $V_\mathrm{WM}$ at different magnetic field values $B_{||}$. The data for $G_\mathrm{LL}$ and $G_\mathrm{LR}$ are shown in the main text. ABSs appear as subgap states in both nonlocal conductances, and are visible in both local conductances $G_\mathrm{LL}$ and $G_\mathrm{RR}$. 

\subsection{Relations between antisymmetric parts of conductance matrix elements}
\label{sec:antisymmetries_dev1}

As discussed in Ref.~\cite{karsten_nl_spectroscopy}, within linear response the antisymmetric parts of conductances $G_\mathrm{LR}$ and $G_\mathrm{LL}$ satisfy the relation
\begin{equation}
\label{eq:anti-symmetry}
    G_\mathrm{LR}^\mathrm{anti}(V_\mathrm{SD}^\mathrm{R}) = -G_\mathrm{LL}^\mathrm{anti}(V_\mathrm{SD}^\mathrm{L})
\end{equation}
as a consequence of current conservation and particle-hole symmetry at voltages below the energy gap of the parent superconductor $e V_\mathrm{SD}^\mathrm{L},\, e V_\mathrm{SD}^\mathrm{R}<\Delta$. Note that this relation compares two quantities that are measured independently in the experiment, as the left-hand side is measured as a function of $V_\mathrm{SD}^\mathrm{R}$ while $V_\mathrm{SD}^\mathrm{L} = \SI{0}{\volt}$ and the right-hand side is measured as a function of $V_\mathrm{SD}^\mathrm{L}$ while $V_\mathrm{SD}^\mathrm{R} = \SI{0}{\volt}$. An equivalent relation between $G_\mathrm{RL}$ and $G_\mathrm{RR}$ exists.\\

\begin{figure}[h]
\includegraphics[scale=0.9]{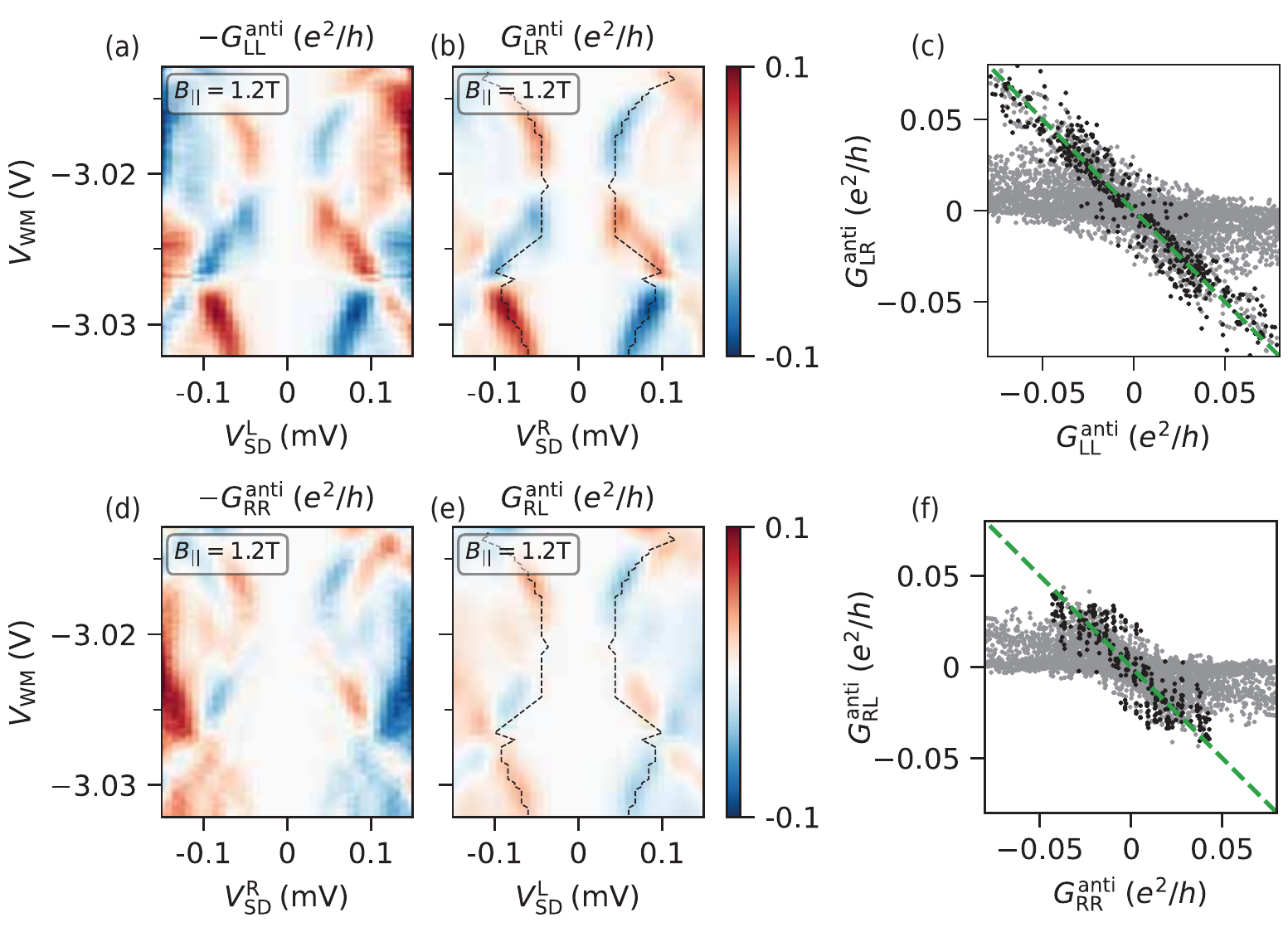}
\caption{\label{fig:antisymmetries_1.2T}(a, b) show $-G_\mathrm{LL}^\mathrm{anti}$ and $G_\mathrm{LR}^\mathrm{anti}$ which are expected to be identical according to a linear transport theory. (c) parametric plot of the data points in (a) and (b). The black data points correspond to the data in (a, b) taken in a three pixel window around the black dashed line in (b). The green dashed line in (c) denotes the relation expected from theory. (d, e) comparison of $-G_\mathrm{RR}^\mathrm{anti}$ and $G_\mathrm{RL}^\mathrm{anti}$. (f) shows a parametric plot of data points from (d, e) with the data points taken from a three pixel window around the dashed line in (e) shown in black.}
\end{figure}

Figure \ref{fig:antisymmetries_1.2T} shows the antisymmetric parts of the local and nonlocal conductances for comparison. The underlying data is the same as in Fig.~\ref{fig:full_plunger_dev_1}(a-d) taken at $B_{||}=\SI{1.2}{\tesla}$. The data in Fig.~\ref{fig:full_plunger_dev_1}(a, b) show $-G_\mathrm{LL}^\mathrm{anti}$ and $G_\mathrm{LR}^\mathrm{anti}$. The two quantities are expected to be the same. The lowest excited state shows up with the same sign and similar strength in both plots. For energy values above the first excited state, the antisymmetric part of the local conductance is larger than the antisymmetric part of the nonlocal conductance. The data points from (a) and (b) are plotted parametrically in (c). The points that correspond to three pixels around the lowest excited state shown by the dashed line in (b) are shown as black dots in (c). They lie close to the green dashed line which corresponds to the linear-response prediction. All other data points are plotted in grey. A large number of these points have a relatively small value in $|G_\mathrm{LR}^\mathrm{anti}|$ compared to their relatively large value in $|G_\mathrm{LL}^\mathrm{anti}|$. These data points originate not only from energies above $\Delta$, where deviations from Eq.~\ref{eq:anti-symmetry} are expected, but also from energies below $\Delta$ and above $\Delta_\mathrm{ind}$.

The quantities $-G_\mathrm{RR}^\mathrm{anti}$ and $G_\mathrm{RL}^\mathrm{anti}$ are plotted in Fig.~\ref{fig:antisymmetries_1.2T}(d, e). A parametric plot of the same data is shown in (f). Figure \ref{fig:antisymmetries_1.2T}(f) also shows that for the lowest energy state data is consistent with Eq.~\ref{eq:anti-symmetry}.\\

\begin{figure}[h]
\includegraphics[scale=0.9]{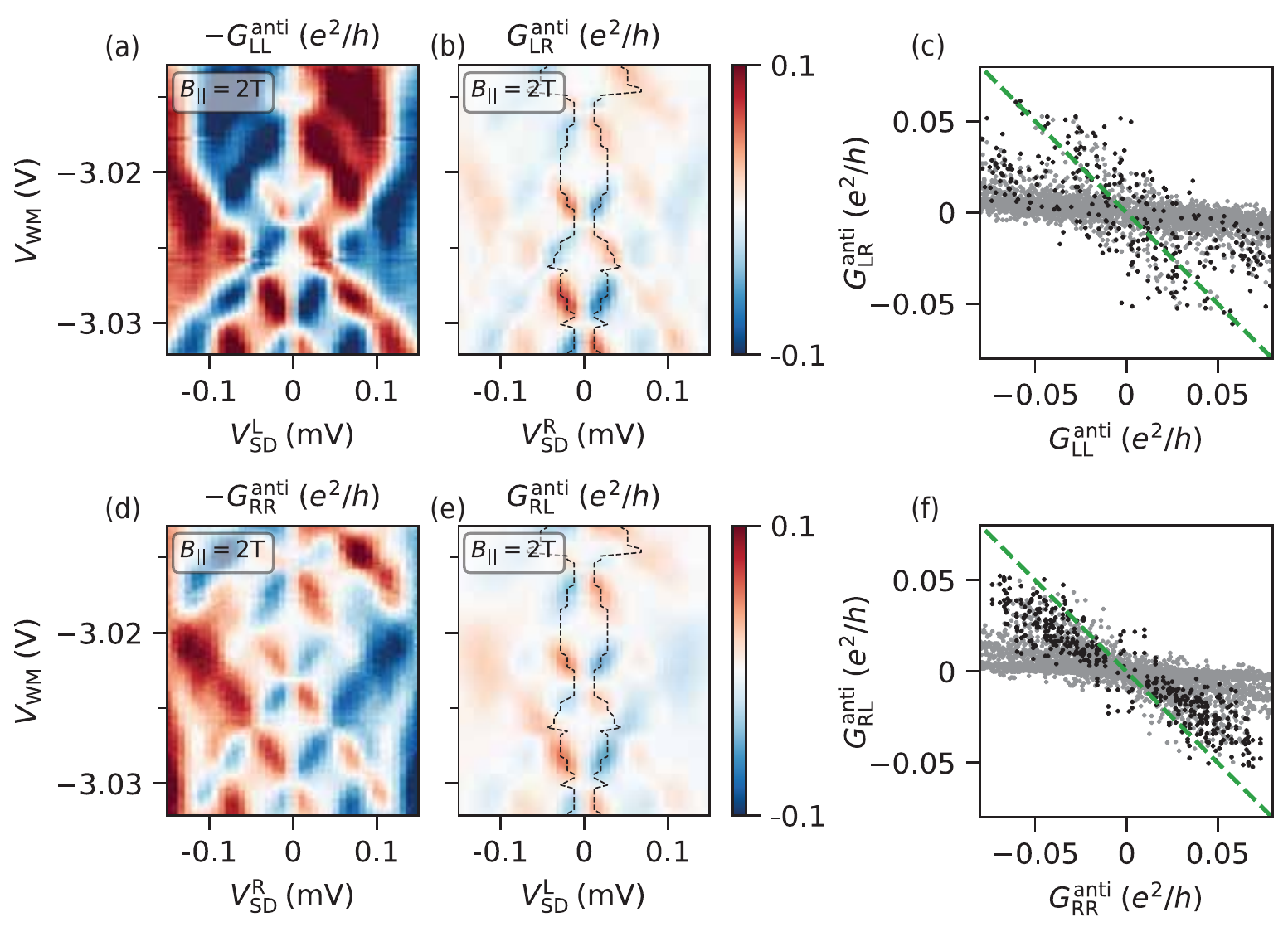}
\caption{\label{fig:antisymmetries_2T}(a, b) show  $-G_\mathrm{LL}^\mathrm{anti}$ and $G_\mathrm{LR}^\mathrm{anti}$ which are expected to be identical according to a linear transport theory. (c) parametric plot of the data points in (a) and (b). The black data points correspond to the data in (a, b) taken in a three pixel window around the black dashed line in (b). The green dashed line in (c) denotes the relation expected from theory. (d, e) comparison of $-G_\mathrm{RR}^\mathrm{anti}$ and $G_\mathrm{RL}^\mathrm{anti}$. (f) shows a parametric plot of data points from (d, e) with the data points taken from a three pixel window around the dashed line in (e) shown in black.}
\end{figure}

For the data taken at $B_{||}=\SI{2}{\tesla}$ in Fig.~\ref{fig:full_plunger_dev_1}(e-h) the same comparison of antisymmetric components of the conductance matrix elements is shown in Fig.~\ref{fig:antisymmetries_2T}. In Fig.~\ref{fig:antisymmetries_2T}(a, b), $-G_\mathrm{LL}^\mathrm{anti}$ and $G_\mathrm{LR}^\mathrm{anti}$ are shown. A parametric plot of the same data is shown in Fig.~\ref{fig:antisymmetries_2T}(c), with the black data points taken in a three pixel window around the lowest energy state denoted by the dashed line in (b). In Fig.~\ref{fig:antisymmetries_2T}(d) and (e), $-G_\mathrm{RR}^\mathrm{anti}$ and $G_\mathrm{RL}^\mathrm{anti}$ are shown for comparison, together with a parametric plot of the same data in Fig.~\ref{fig:antisymmetries_2T}(f). The  low-energy state shows up with the same sign in $-G^\mathrm{anti}_\mathrm{LL}$ and $G_\mathrm{LR}^\mathrm{anti}$. The values of $-G^\mathrm{anti}_\mathrm{RR}$ and $G_\mathrm{RL}^\mathrm{anti}$ for the low-energy state are similar. The parametric plots in Fig.~\ref{fig:antisymmetries_2T}(f) demonstrate that the data taken around the first excited state lie close to dashed green line, which indicates the linear-response relation for $eV_\mathrm{SD}<\Delta$. The black points in Fig.~\ref{fig:antisymmetries_2T}(c) stemming from the low-energy state appear less correlated according to the theory expectation compared to the lower field value $B_{||}=\SI{1.2}{\tesla}$.\\

The relation between the antisymmetric parts of the conductance matrix elements in Eq.~\ref{eq:anti-symmetry} has as a consequence that the sum of all conductance matrix elements 
\begin{equation}
    G_\mathrm{sum} (V_\mathrm{SD})= G_\mathrm{LL} (V_\mathrm{SD}^\mathrm{L})+G_\mathrm{RL}(V_\mathrm{SD}^\mathrm{L})+G_\mathrm{LR}(V_\mathrm{SD}^\mathrm{R})+G_\mathrm{RR}(V_\mathrm{SD}^\mathrm{R})
\end{equation}
is a symmetric function of $V_\mathrm{SD}$ \cite{karsten_nl_spectroscopy}. 

\begin{figure}[h]
\includegraphics[scale=0.9]{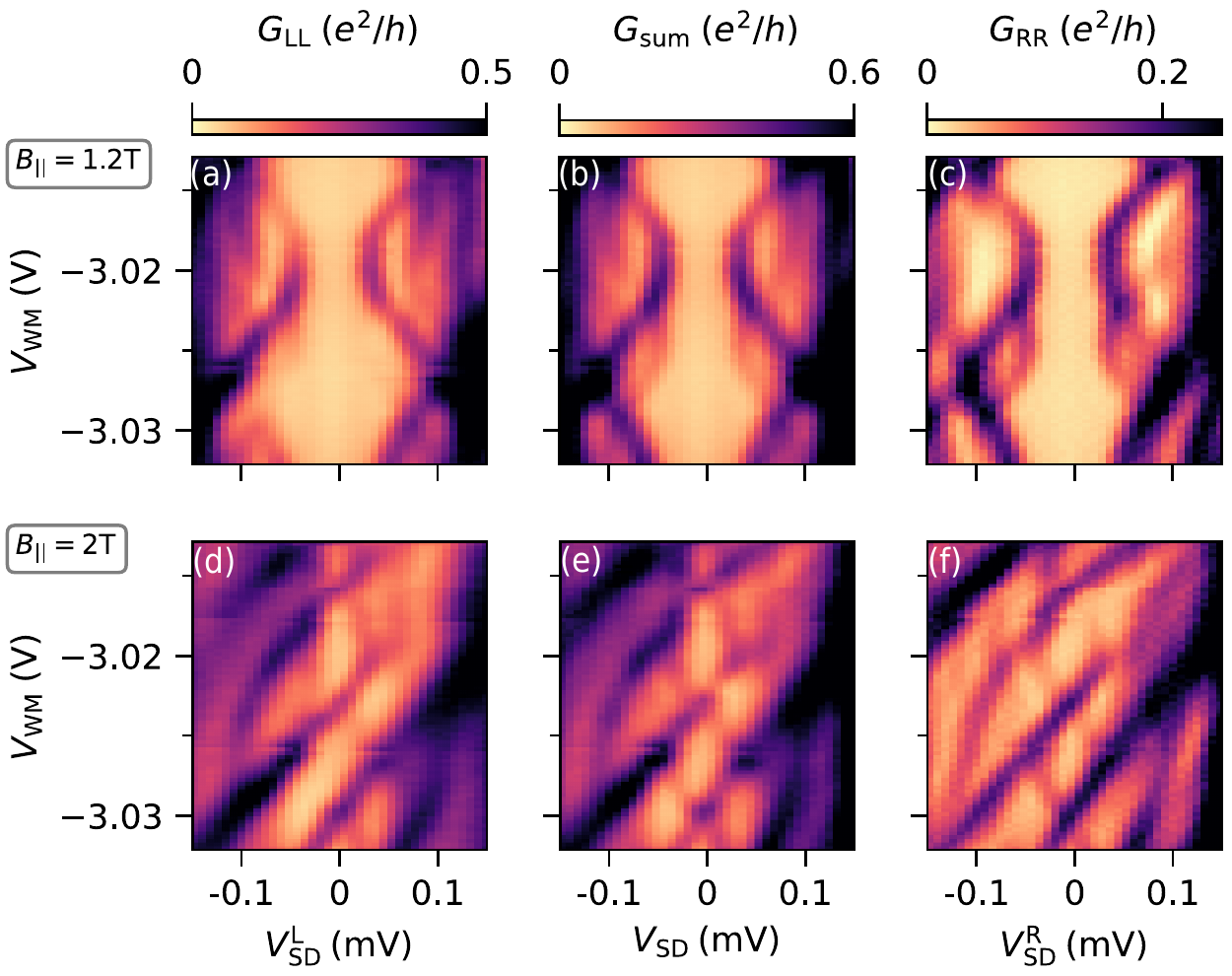}
\caption{\label{fig:symmetries_1.2T}(a) Tunneling conductance $G_\mathrm{LL}$, (b) the sum of all conductance matrix elements $G_\mathrm{sum}$, (c) tunneling conductance $G_\mathrm{RR}$ measured at $B_{||}=\SI{1.2}{\tesla}$. Subgap states around $V_\mathrm{WM} = \SI{-3.030}{\volt}$ and $V_\mathrm{WM}=\SI{-3.015}{\volt}$ appear symmetric with respect to $V_\mathrm{SD}$ in $G_\mathrm{sum}$, but show a notable asymmetry in $G_\mathrm{LL}$ and $G_\mathrm{RR}$. (d) tunneling conductance $G_\mathrm{LL}$, (e) the sum of all conductance matrix elements $G_\mathrm{sum}$, (f) tunneling conductance $G_\mathrm{RR}$ measured at $B_{||}=\SI{2}{\tesla}$. The low-energy state appears symmetric with respect to $V_\mathrm{SD}$ in $G_\mathrm{sum}$.}
\end{figure}

The quantity $G_\mathrm{sum}$ calculated from the data in Fig.~\ref{fig:full_plunger_dev_1}(a-d) measured at $B_{||}=\SI{1.2}{\tesla}$ is shown in Fig.~\ref{fig:symmetries_1.2T}(b). The two corresponding local conductances $G_\mathrm{LL}$ and $G_\mathrm{RR}$ are plotted for comparison in  Fig.~\ref{fig:symmetries_1.2T}(a, c). Qualitatively, $G_\mathrm{sum}$ is more symmetric in comparison to $G_\mathrm{LL}$  and $G_\mathrm{RR}$ for some of the subgap states.\\

$G_\mathrm{sum}$ calculated for data taken at $B_{||}=\SI{2}{\tesla}$ in Fig.~\ref{fig:full_plunger_dev_1}(i-l) is shown in Fig.~\ref{fig:symmetries_1.2T}(e). The two local conductances $G_\mathrm{LL}$  and $G_\mathrm{RR}$ are plotted in Fig.~\ref{fig:symmetries_1.2T}(d, f) for comparison. $G_\mathrm{sum}$ appears more symmetric with respect to source drain voltage $V_\mathrm{SD}$. In particular, the low-energy state that oscillates around zero bias appears more symmetric in $G_\mathrm{sum}$ compared to $G_\mathrm{LL}$ and $G_\mathrm{RR}$. States at larger source drain voltages appear with a strong anti-symmetry in $G_\mathrm{sum}$ which is a result of the relation in Eq.~\ref{eq:anti-symmetry} only being fulfilled for the low-energy state in the experiment.\\

\subsection{Extracted value of $Q_\mathrm{L}$ and $Q_\mathrm{R}$}

For the nonlocal conductances measured at \SI{1.2}{\tesla} shown in Fig.~\ref{fig:full_plunger_dev_1}(b, d) we extracted
\begin{equation}
\label{eq:Q}
Q_\mathrm{R}=\mathrm{sign}(V_\mathrm{SD}^\mathrm{R}) \frac{G_\mathrm{LR}^\mathrm{sym}(V_\mathrm{SD}^\mathrm{R})}{G_\mathrm{LR}^\mathrm{anti}(V_\mathrm{SD}^\mathrm{R})}
\end{equation}
for the lowest energy ABS at $E=e V_\mathrm{SD}^\mathrm{R}$. This quantity is expected to reflect the local charge character of the state at the location of the right tunnel probe. The analogously defined quantity $Q_\mathrm{L}$ is a measure of the local charge character at the location of the left tunnel probe.

\begin{figure}[h]
\includegraphics[scale=0.9]{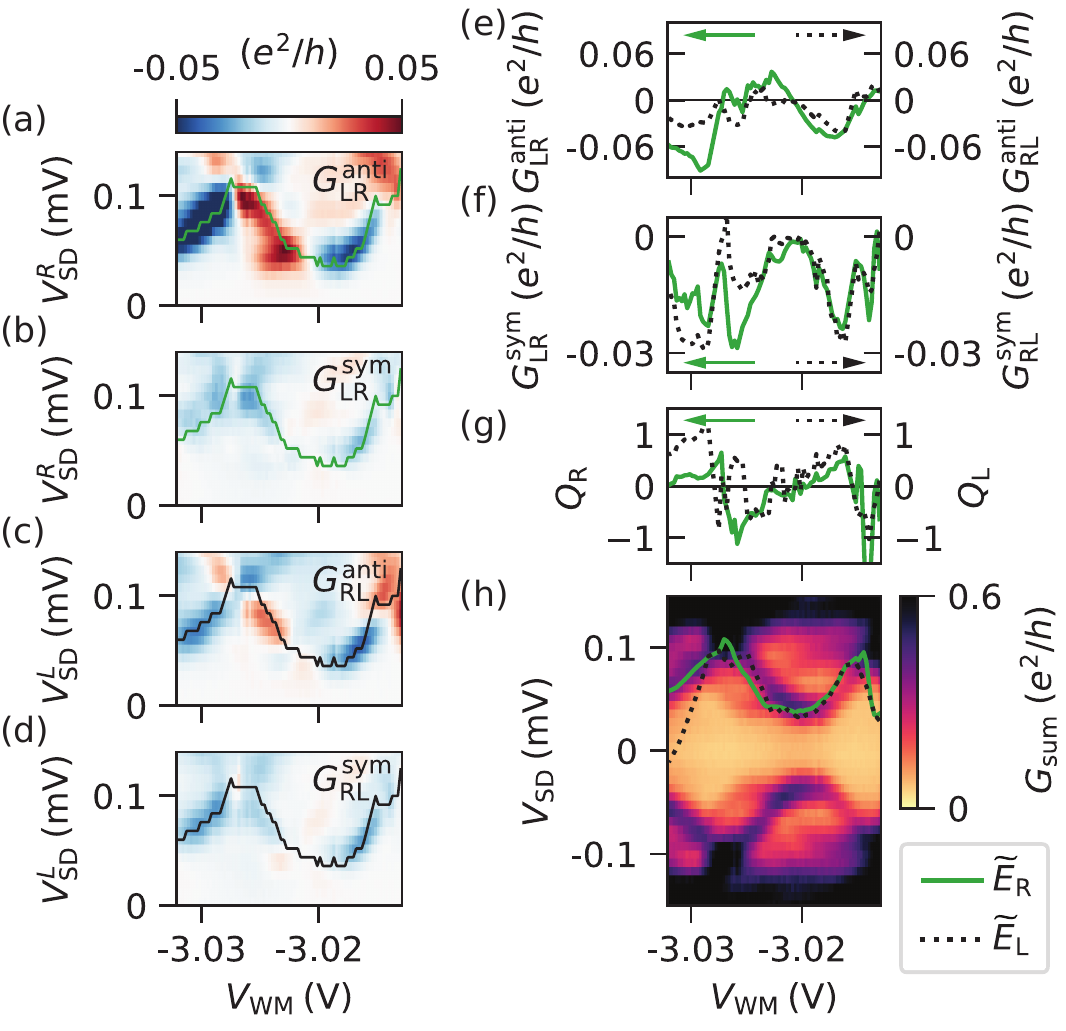}
\caption{\label{fig:Q_extraction_left_1.2T}(a-d) Symmetric and antisymmetric components of the nonlocal conductances measured at $B_{||}=\SI{1.2}{\tesla}$. The values at the position of the green and black line are plotted in (e, f) and were used to extract $Q_\mathrm{L}$, $Q_\mathrm{R}$, which are shown in (g). The two evolve similarly with $V_\mathrm{WM}$. An abrupt change from positive to negative values is seen where neighboring ABSs cross. (h) shows $G_\mathrm{sum}$, the sum of the four conductance components. The dotted black, solid green lines display the energies $\widetilde{E}_\mathrm{L}$, and $\widetilde{E}_\mathrm{R}$ inferred from the integrated $Q_\mathrm{L}$, and $Q_\mathrm{R}$. They roughly retrace the bound state energy.}
\end{figure}

Figure \ref{fig:Q_extraction_left_1.2T}(a-d) shows the symmetric and antisymmetric components of the nonlocal conductances. The values at the position of the lowest excited state are extracted at the position given by the green and black lines. The resulting values are plotted in Fig.~\ref{fig:Q_extraction_left_1.2T}(e, f). The resulting values for $Q_\mathrm{L}$ and $Q_\mathrm{R}$ are shown in Fig.~\ref{fig:Q_extraction_left_1.2T}(g). The two quantities show roughly the same evolution as a function of gate voltage $V_\mathrm{WM}$. At the two points where ABSs cross around $V_\mathrm{SD}=\SI{0.1}{\milli\volt}$ an abrupt change from positive to negative is observed in both $Q_\mathrm{L}$ and $Q_\mathrm{R}$. Around the gate voltage, where the ABS goes through a minimum in energy, a smooth change from negative to positive $Q_\mathrm{L}$ and $Q_\mathrm{R}$  can be seen. Regions of positive and negative valued quasiparticle charge coincide with ranges where the subgap state has a positive or negative slope as a function of $V_\mathrm{WM}$ respectively.  Integrating $Q_\mathrm{j}$ ($j\in\{\mathrm{L},\:\mathrm{R}\}$), rescaling, and taking into account a linear background leads to the inferred energies
\begin{equation}\label{eq:Einf}
    \begin{aligned}
    \widetilde{E}_\mathrm{j}=a\int Q_{j} \mathrm{d}V_\mathrm{WM} + b V_\mathrm{WM} + c.
    \end{aligned}
\end{equation} 
$a$, $b$, and $c$ are free parameters.
The resulting $\widetilde{E}_\mathrm{L}$ and $\widetilde{E}_\mathrm{R}$   are plotted in Fig.~\ref{fig:Q_extraction_left_1.2T}(h). They approximately retrace the energy of the lowest lying subgap state.
This evolution of $Q_\mathrm{L}\approx Q_\mathrm{R}$ in accordance with the energy of the subgap state suggests that $Q_\mathrm{L}$ and $Q_\mathrm{R}$ measure the total charge of the subgap state. \\

\begin{figure}[h]
\includegraphics[scale=0.9]{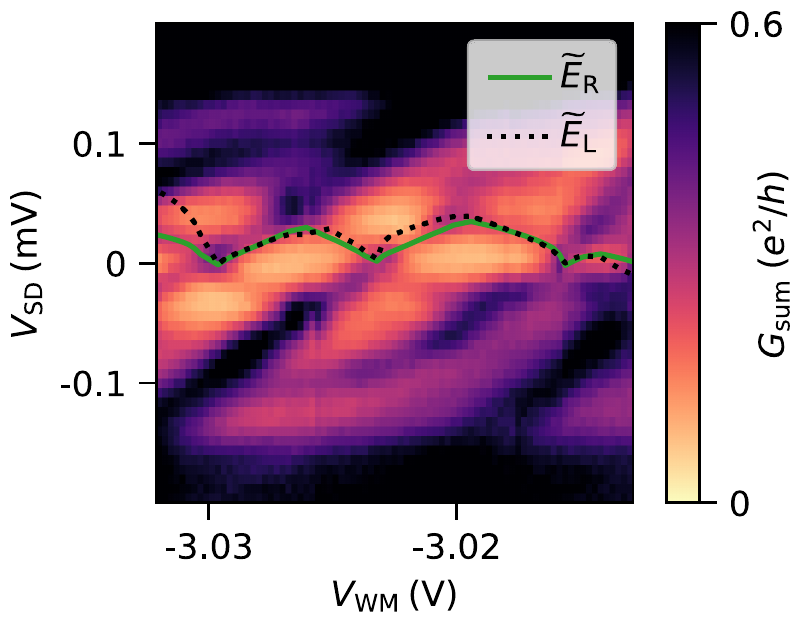}
\caption{\label{fig:E_inf_2T}Map of the sum of all conductances $G_\mathrm{sum}$ at $B_{||}=\SI{2}{\tesla}$. The inferred energies $\widetilde{E}_\mathrm{L}$, and $\widetilde{E}_\mathrm{R}$ from the integrated $Q_\mathrm{L}$, and $Q_\mathrm{R}$ are superimposed. The $\widetilde{E}_\mathrm{L}$, and $\widetilde{E}_\mathrm{R}$ curves roughly retrace the energy of the subgap state.}
\end{figure}

For the values of $Q_{L}$, $Q_{R}$ at $B_{||}=\SI{2}{\tesla}$ presented in the main part, the inferred energy curves $\widetilde{E}_\mathrm{L}$ and $\widetilde{E}_\mathrm{R}$   are plotted in Fig.~\ref{fig:E_inf_2T} superimposed on a map showing the sum of all conductances $G_\mathrm{sum}$. The curves roughly match the evolution of the bound state energy which indicates that the experimentally extracted values for $Q_{L}$, $Q_{R}$ measure the total, integrated charge of the bound state \cite{karsten_nl_spectroscopy}. The values of the free parameters $a$, $b$, $c$ that were used in the individual figures in this work are summarized in Table \ref{table:abc_param}.

\begin{table}[h!]
\begin{center}
\begin{tabular}{ |p{1.5cm}||p{2cm}|p{2cm}|p{2cm}|p{2cm}|p{2cm}|p{2cm}|  }
 \hline
 & \multicolumn{3}{c|}{$\widetilde{E}_\mathrm{L}$} & \multicolumn{3}{c|}{$\widetilde{E}_\mathrm{R}$}  \\
 \hline
 Figure& a [$\mathrm{meV/V}$] & b [$\mathrm{meV/V}$] & c [$\mathrm{meV}$] &a [$\mathrm{meV/V}$] & b [$\mathrm{meV/V}$] & c [$\mathrm{meV}$]  \\
 \hline 
 Fig.~\ref{fig:Q_extraction_left_1.2T}(h) & 35.0 & 5.0 & 0.03 & 35.0 & -3.0 & 0.03\\
 \hline
 Fig.~\ref{fig:E_inf_2T} & 7.0 & 1.7 & -0.01 & 8.5 & -1.5 & 0.00\\
 \hline
 Fig.~\ref{fig:Q_extr_right_1.2T}(h) & 12.0 & -4.0 & 0.05 & 12.0 & -6.0 & 0.07\\
 \hline
 Fig.~\ref{fig:Q_extr_dev_2_0.2T}(h) & 12.0 & 0.0 & 0.10 & 12.0 & 0.0 & 0.10\\
 \hline
 Fig.~\ref{fig:Q_extr_dev_2_1.7T}(h) & 9.0 & 7.0 & -0.25 & 9.0 & 2.0 & -0.15\\
 \hline
\end{tabular}
\end{center}
\caption{Free parameters $a$, $b$, $c$ used to match $\widetilde{E}_\mathrm{L}$ and $\widetilde{E}_\mathrm{R}$ to the energy of the low-energy ABS as a function gate voltage.}
\label{table:abc_param}
\end{table}

\subsection{Local and nonlocal conductance as a function of out-of-plane magnetic field $B_{\perp}$}

\begin{figure}[h]
\includegraphics[scale=0.9]{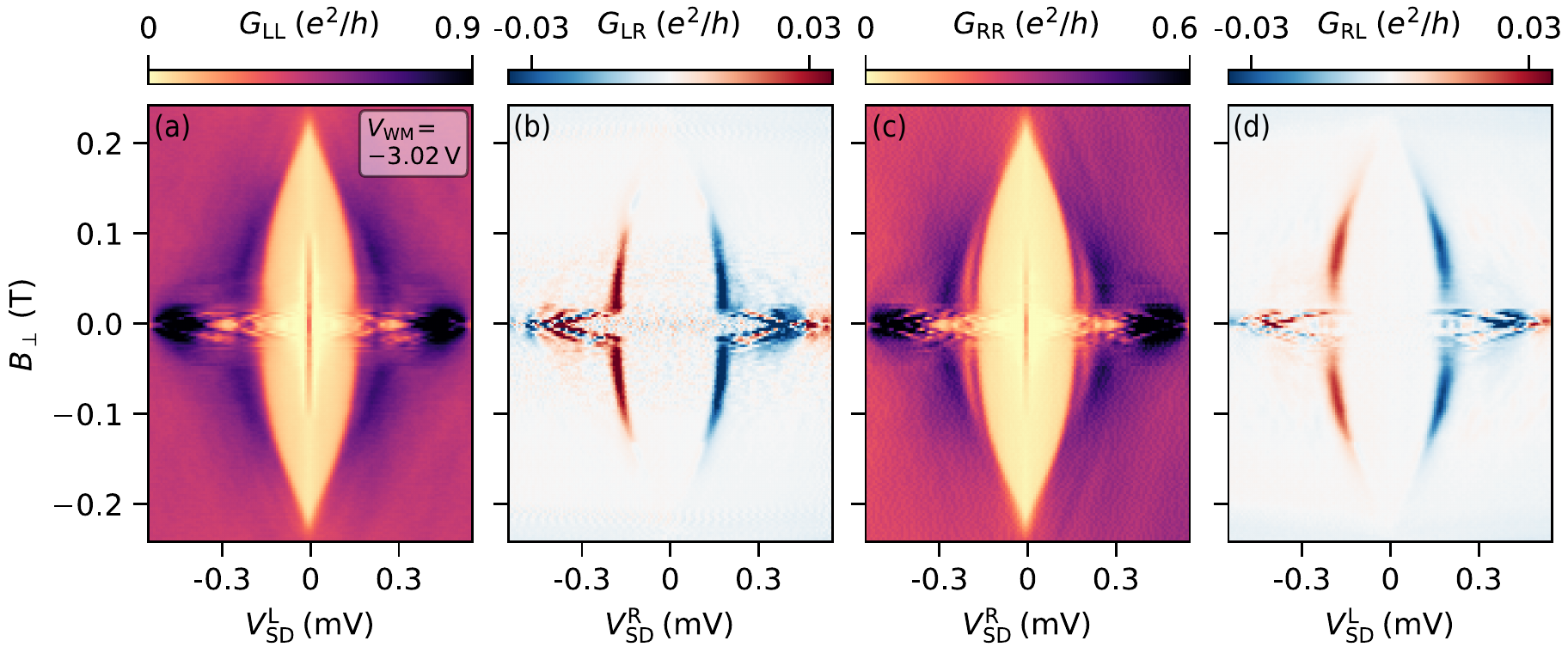}
\caption{\label{fig:perp_fieldscan_left}Measurement of all four conductances as a function of magnetic field $B_{\perp}$ pointing out of the 2DEG plane. (a, c) The local conductances show a clear supercurrent as a peak at zero source drain bias. The supercurrent does not appear in the nonlocal conductances in (b, d).}
\end{figure}

Figure \ref{fig:perp_fieldscan_left} shows local and nonlocal conductance as a function of magnetic field $B_\perp$ applied perpendicular to the 2DEG plane. The direction of the magnetic field is denoted in Fig.~\ref{fig:detection}. The critical field value of the Al film for this magnetic field direction is lowered in comparison to the in plane magnetic field $B_{||}$. The superconducting gap therefore closes at a magnetic field value around \SI{0.22}{\tesla} before subgap states enter the superconducting energy gap. The nonlocal conductance is restricted to a small region close to the gap of the superconductor. 

A supercurrent is visible at $V_\mathrm{SD}=\SI{0}{\volt}$ for small perpendicular fields, $|B_{\perp}|<\SI{0.1}{\tesla}$. In this regime, both the wire and the leads are superconducting. The supercurrent appears clearly in the local conductances $G_\mathrm{LL}$ and $G_\mathrm{RR}$, but does not appear in the nonlocal conductances $G_\mathrm{LR}$ and $G_\mathrm{RL}$. This is a result of the supercurrent being mediated by Cooper pairs that go from the superconducting leads directly to the superconducting condensate of the NW. Nonlocal conductance, on the other hand, is mediated by $1e$ quasiparticles that are transmitted and Andreev reflected from one lead to the other.

\section{Nonlocal conductance spectroscopy of Andreev bound state localized under gate $\mathrm{W_R}$}

The multigate design of device 1, shown in Fig.~\ref{fig:detection},  allows for both nonlocal spectroscopy of an ABS under the gate $\mathrm{W_M}$ as demonstrated in the main text and measurements of an ABS under the gate $\mathrm{W_R}$. For this purpose, current to voltage converting amplifiers were connected to the leads under the gates $\mathrm{T_R}$ and $\mathrm{T_S}$, while the lead under the gate $\mathrm{T_L}$ was terminated with an open circuit at the breakout box. 

In the following, data on an ABS under the gate $\mathrm{W_R}$ are shown. The tunnel probe under gate $\mathrm{T_R}$ and the respective current is labeled $j=L$. The tunnel probe under $\mathrm{T_S}$ and the respective current is labeled $j=R$. This follows the commonly applied naming convention for nonlocal conductance, i.e., $G_\mathrm{LL}$ and $G_\mathrm{LR}$ are a result of a tunneling current at the left side of the NW segment of interest and $G_\mathrm{RR}$ and $G_\mathrm{RL}$ are a result of a tunneling current at the right side of the NW segment of interest.

\subsection{Measurements of the conductance matrix}

In order to measure an ABS that is confined to the NW segment under the gate $\mathrm{W_{R}}$, the gate voltages $V_\mathrm{WL},\; V_\mathrm{WM},\; V_\mathrm{WS}$ were set to $\SI{-7}{\volt}$. A measurement of the full conductance matrix as a function of magnetic field $B_{||}$ is shown in Fig.~\ref{fig:fieldscan_dev_2} with the gate voltage $V_\mathrm{WR}=\SI{-3.09}{\volt}$ - significantly more positive than the neighboring gates. The ABS appears as subgap states in both local conductances $G_\mathrm{LL}$ and $G_\mathrm{RR}$. The states emerge from the continuum of quasiparticles at an energy $\Delta$ at low magnetic field $B_{||}\approx\SI{0.3}{\tesla}$ and cross zero bias at $B_{||}\approx\SI{1.6}{\tesla}$. The nonlocal conductances $G_\mathrm{LR}$ and $G_\mathrm{RL}$ are appreciably large at the energies of the low energy ABS and above. The signal is furthermore strongly suppressed for source drain voltages above the parent gap $eV_\mathrm{SD}>\Delta$. The nonlocal conductance shows a strong antisymmetric component and the subgap states originating from the low energy ABS show a characteristic pattern as they cross zero energy. Specifically, the states cross without a change in sign of the nonlocal conductance for the state going from negative to positive values of $V_\mathrm{SD}$. Similar behavior has been observed previously in numerical studies and experiments \cite{karsten_nl_spectroscopy,gerbold_nonlocal,SDS_nl_conductance}.\\

\begin{figure}[h]
\includegraphics[scale=0.9]{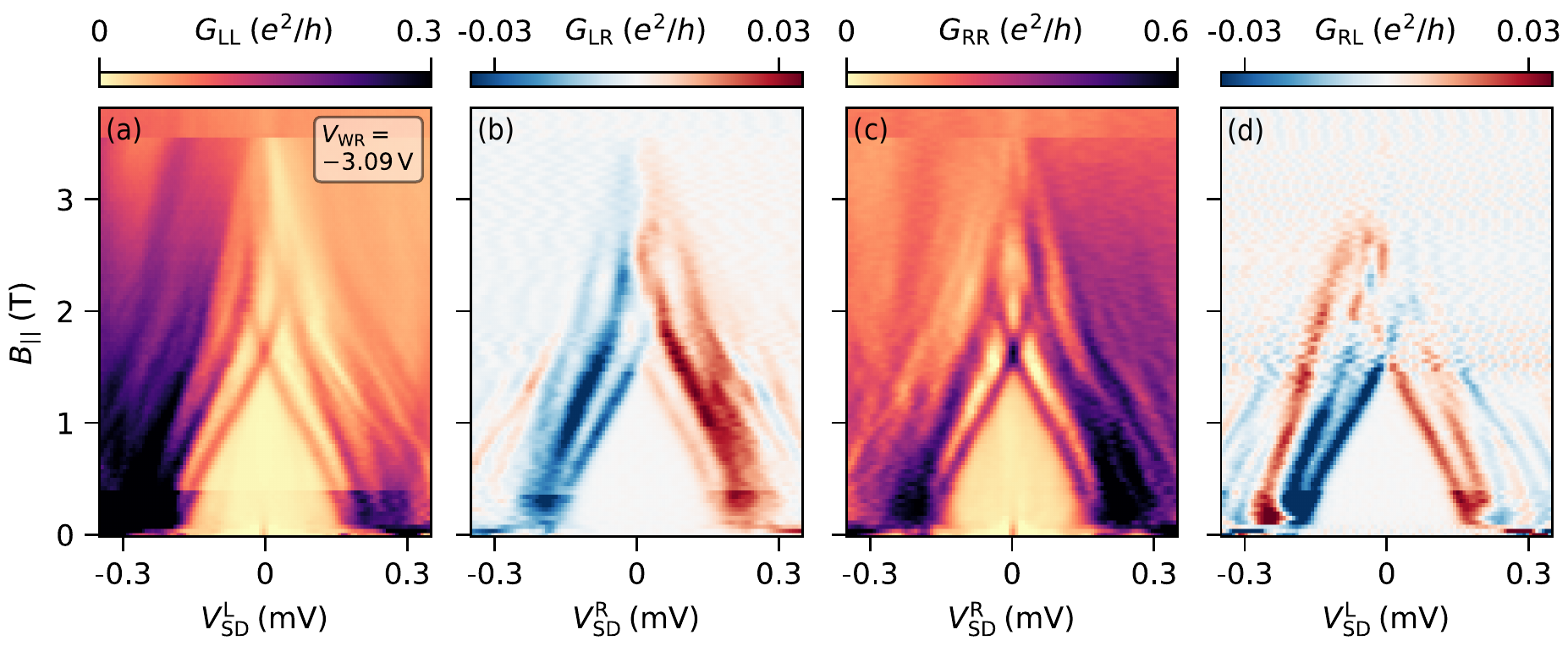}
\caption{\label{fig:fieldscan _dev_1_right}Measurement of all four conductances as a function of magnetic field $B_{||}$. (a, c) Local conductances reveal an ABS in the form of subgap states that cross zero source drain bias. (b, d) ABSs and higher excited states appear with sizeable nonlocal conductance.}
\end{figure}

\begin{figure}[h]
\includegraphics[scale=0.9]{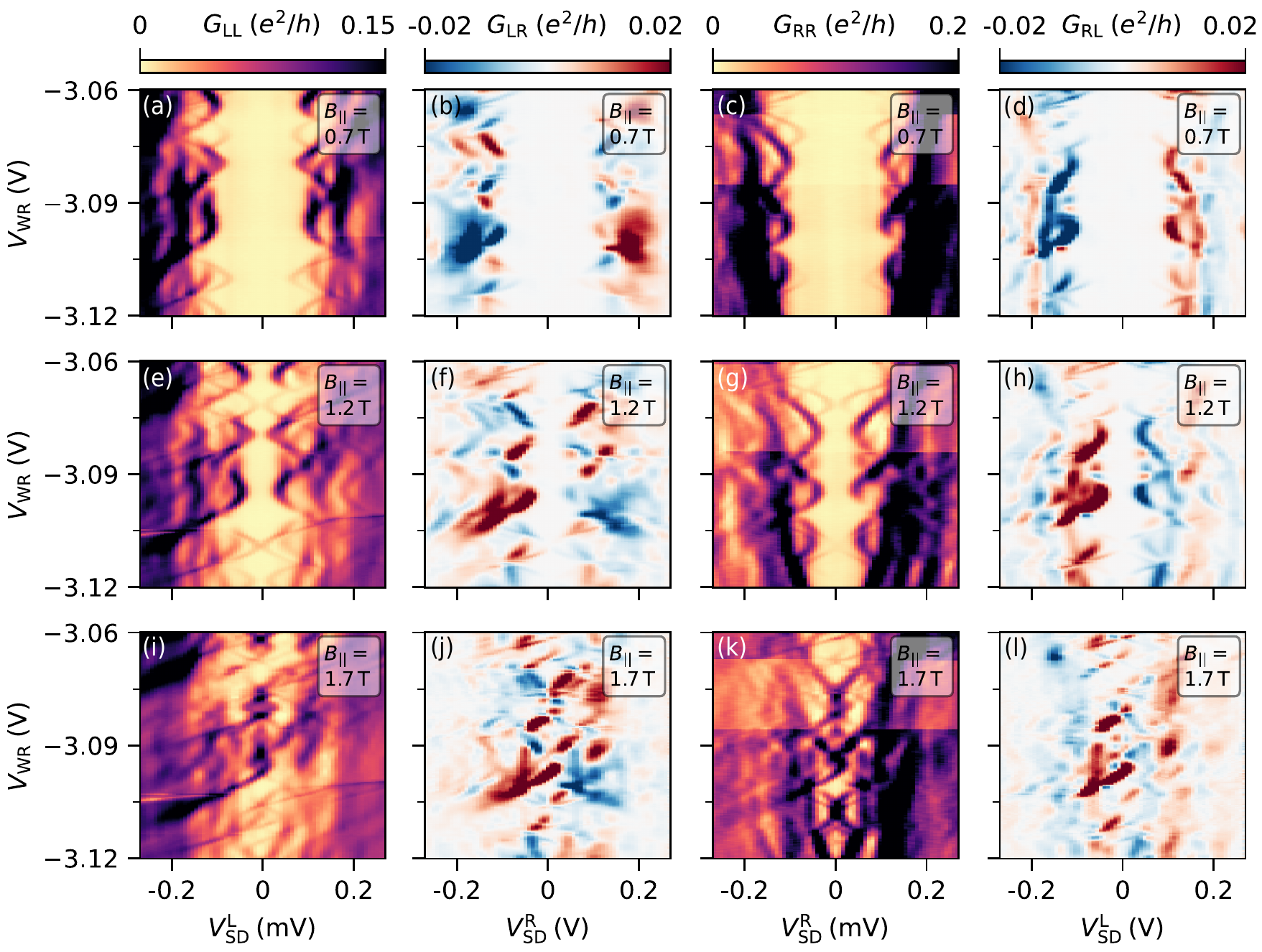}
\caption{\label{fig:plungerscans_dev_1_right}Each row shows all four conductances as a function of gate voltage $V_\mathrm{WR}$ at a specific value of $B_{||}$. (a-d) ABSs in the NW appear as lobe shaped subgap states that appear in both local conductances and nonlocal conductances at low magnetic fields. (e-h) With increasing filed $B_{||}$ the ABS are lowered in energy. (i-l) The subgap states merged and cross at zero bias.}
\end{figure}

The evolution of conductances as a function of gate voltage $V_\mathrm{WR}$ is shown in Fig.~\ref{fig:plungerscans_dev_1_right}. Note that the voltages on the gates $\mathrm{T_R}$ and $\mathrm{T_S}$ were compensated according to the equations

\begin{equation}
 \begin{aligned} \label{eq:long_plunger}
     V_\mathrm{TR}&=-\SI{0.01}{\volt}-\frac{0.05}{0.40} \cdot(V_\mathrm{WR}+\SI{3.1}{\volt})\\
     V_\mathrm{TS}&=-\SI{0.14}{\volt}-\frac{0.045}{0.40} \cdot (V_\mathrm{WR}+\SI{3.5}{\volt}) 
 \end{aligned}
\end{equation}

in order to keep the transparency of the tunnel barriers constant throughout the measurement. To denote that more than one gate voltage was changed, the vertical axis in \ref{fig:plungerscans_dev_1_right} is denoted $\tilde{V}_\mathrm{WR}$ instead of ${V}_\mathrm{WR}$.

Data taken at a magnetic field $B_{||}=\SI{0.7}{\tesla}$ in \ref{fig:plungerscans_dev_1_right}(a-d) show subgap states as lobes close to the edge of the parent gap $\Delta$. The states appear in both local conductances $G_\mathrm{LL}$ and $G_\mathrm{RR}$. In the nonlocal conductances $G_\mathrm{LR}$ and $G_\mathrm{RL}$ subgap states can be seen with the charactereistic change in sign at points of minimal energy of the ABS or at intersection points of ABSs.

At a magnetic field value $B_{||}=\SI{1.2}{\tesla}$ in \ref{fig:plungerscans_dev_1_right}(e-h) the ABSs are lowered in energy due to an increased Zeeman energy. Between the low-energy state and the parent gap $\Delta\approx\SI{0.2}{\milli\eV}$, higher excited states appear both in the local and nonlocal conductances. Characteristic sign changes are observed in nonlocal conductance at crossing points of ABSs at finite bias $V_\mathrm{SD}\approx\SI{0.1}{\milli\volt} $and at gate voltages at which the ABSs reach a minimum in energy.

The ABSs reach zero bias at a magnetic field value of $B_{||}=\SI{1.7}{\tesla}$ (see Fig.~\ref{fig:fieldscan _dev_1_right}). This results in a dense spectrum of subgap states that fill the superconducting gap in both the local and nonlocal conductances measured in \ref{fig:plungerscans_dev_1_right}(i-l). \\

\subsection{Extracted value of $Q_\mathrm{L}$ and $Q_\mathrm{R}$}

\begin{figure}[h]
\includegraphics[scale=0.85]{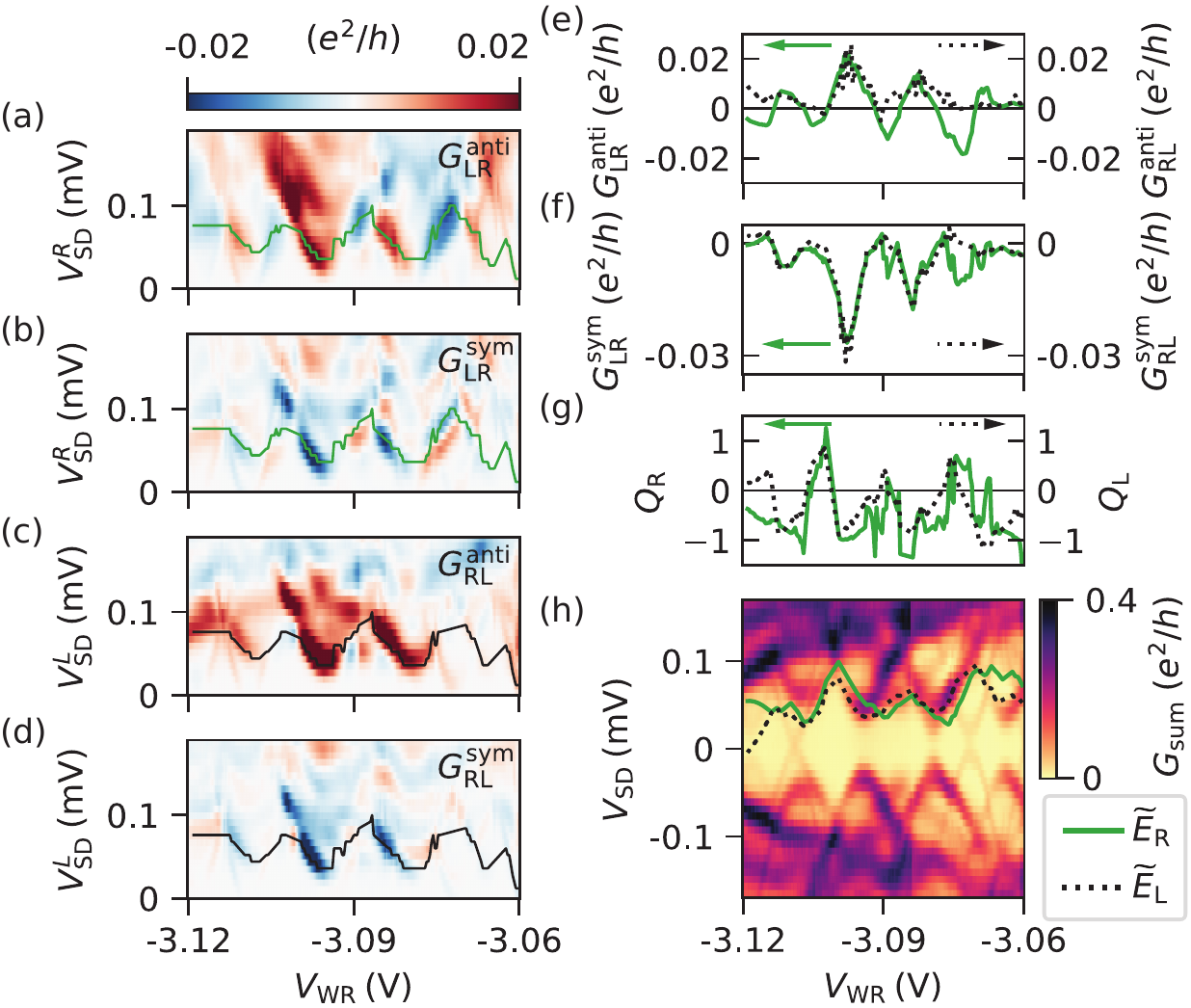}
\caption{\label{fig:Q_extr_right_1.2T}(a-d) Symmetric and antisymmetric components of the nonlocal conductances measured at $B_{||}=\SI{1.2}{\tesla}$. The values at the position of the green and black lines are plotted in (e, f) and were used to calculate $Q_\mathrm{L}$, $Q_\mathrm{R}$, which are plotted in (g). The two quantities evolve similarly with $V_\mathrm{WM}$. (h) shows the sum of all conductances $G_\mathrm{sum}$. The dotted black, solid green lines display the energies $\widetilde{E}_\mathrm{L}$, $\widetilde{E}_\mathrm{R}$ inferred from the integrated $Q_\mathrm{L}$, $Q_\mathrm{R}$.}
\end{figure}

Figure \ref{fig:Q_extr_right_1.2T}(a-d) shows the symmetric and antisymmetric components of $G_\mathrm{LR}$ and $G_\mathrm{RL}$ measured at $B_{||}=\SI{1.2}{\tesla}$. The values at the position of the lowest lying state [shown by the green and black line in Fig.~\ref{fig:Q_extr_right_1.2T}(a-d)] are plotted in Fig.~\ref{fig:Q_extr_right_1.2T}(e, f). The values of $Q_\mathrm{L}$ and $Q_\mathrm{R}$ plotted in Fig.~\ref{fig:Q_extr_right_1.2T}(g) are similar. The energies $\widetilde{E}_\mathrm{L}$ and $\widetilde{E}_\mathrm{R}$ inferred from the integration of $Q_\mathrm{L}$ and $Q_\mathrm{R}$ according to Eq.~\ref{eq:Einf} track the energy of the lowest lying state for a finite range of $V_\mathrm{WM}$ as seen from Fig.~\ref{fig:Q_extr_right_1.2T}(h). \\

\section{Data on device 2}

\begin{figure}[h]
\includegraphics[scale=0.9]{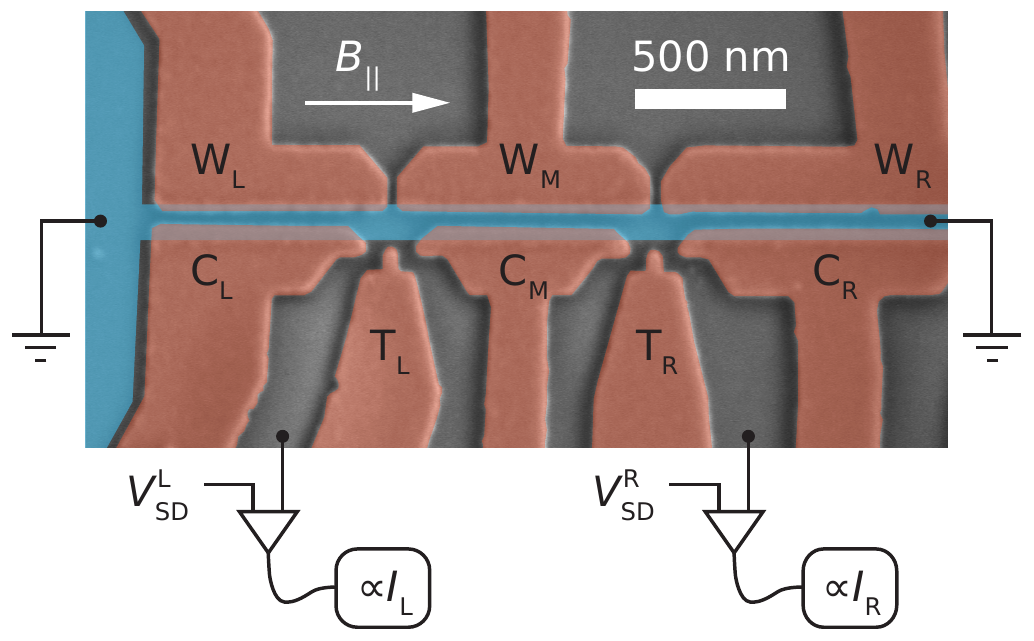}
\caption{\label{fig:dev_2_sem}False-color micrograph of device 2. A wire connected to electrical ground planes at its ends is formed by the Al (blue) on top of the InAs quantum well (gray). Electrostatic gates (red) control the electron density in the NW and control the quantum point contacts adjacent to the NW.}
\end{figure}

In the following, we present data on device 2, which is shown in the false-color electron micrograph in Fig.~\ref{fig:dev_2_sem}. The device consists of an Al film (shown in blue) that is shaped into a NW connected to ground planes at both of its ends. A single layer of electrostatic gates made from Ti/Au (shown in red) on top of $\mathrm{HfO_x}$ gate dielectric are used to control the device and shape the electron density in the 2DEG (shown in gray). The gates labeled $\mathrm{W_L}$, $\mathrm{W_M}$, $\mathrm{W_R}$ are used to electrostatically confine the NW and tune the electron density in the semiconductor under the Al. The gates labeled $\mathrm{C_L}$, $\mathrm{C_M}$, $\mathrm{C_R}$ also electrostatically confine the NW under the Al, and in addition form quantum point contacts adjacent to the NW. The gates $\mathrm{T_L}$ and $\mathrm{T_R}$ offer additional control over the tunnel barrier formed by the point contact. The gates $\mathrm{T_L}$ and $\mathrm{T_R}$ screen the 2DEG region underneath them, which serve as semiconducting leads for tunneling measurements. Device 2 is based on an $\mathrm{In_{1-y}Al_{y}As-InAs-In_{1-x}Ga_{x}As}$ quantum well. Due to the larger band gap of $\mathrm{In_{1-y}Al_{y}As}$ compared to $\mathrm{In_{1-x}Ga_{x}As}$, the interface transparency between the InAs and superconducting Al is decreased for device 2 in comparison to device 1. \\

Device 2 used the same measurement setup as device 1 (see: Fig.~\ref{fig:detection}). Each of the ground planes of Al at the NW ends are connected by two electrical lines and grounded at the breakout box. Each of the semiconducting leads under the gates $\mathrm{T_L}$ and $\mathrm{T_R}$ are connected to a current to voltage converting amplifier in order to measure the tunneling currents $I_\mathrm{L}$ and $I_\mathrm{R}$. Using the lock-in detection scheme as shown in Fig.~\ref{fig:detection} allows for the measurement of all four conductances $G_\mathrm{ij}=\mathrm{d}I_\mathrm{i}/\mathrm{d}V_\mathrm{SD}^\mathrm{j}$ ($i,j\in \{\mathrm{L}, \mathrm{R}\}$).\\

\subsection{Measurement of local and nonlocal conductance: Magnetic field dependence}

\begin{figure}[h]
\includegraphics[scale=0.9]{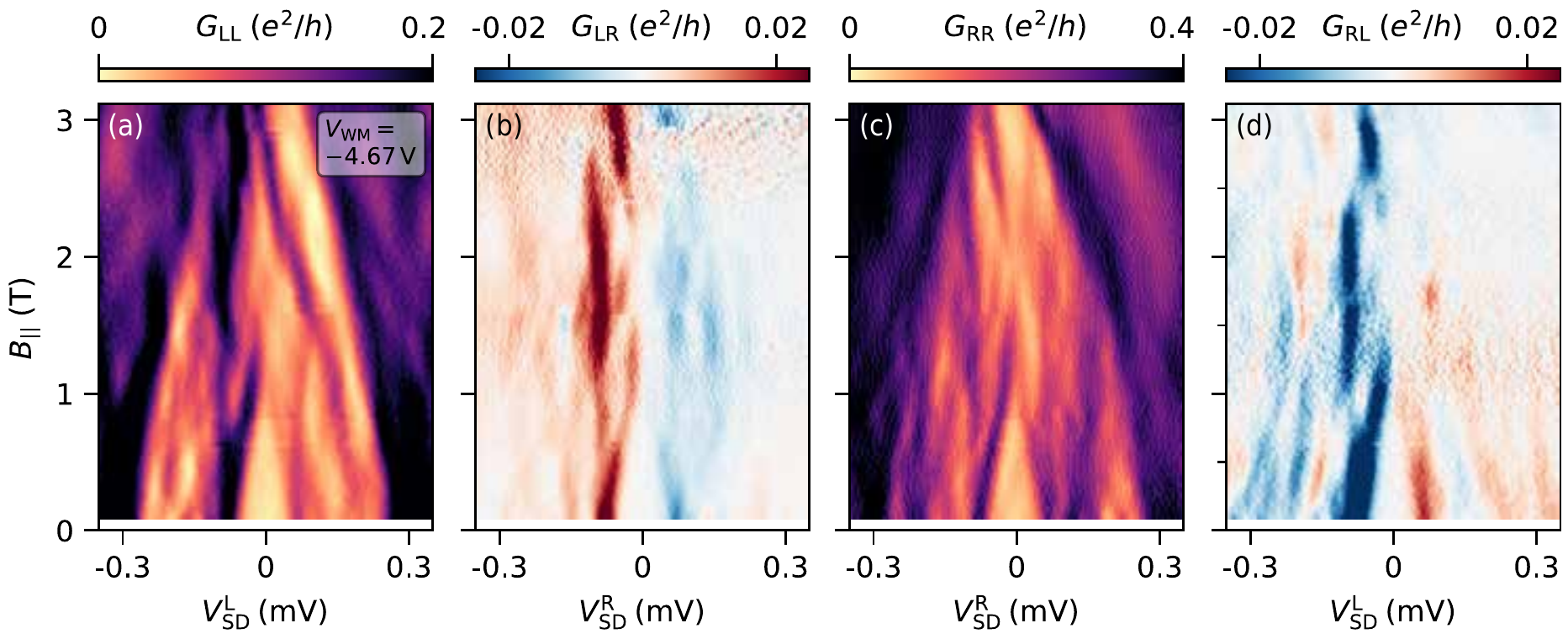}
\caption{\label{fig:fieldscan_dev_2}Local and nonlocal conductances as a function of magnetic field measured on device 2. The parent gap $\Delta$ is visible at high source drain bias in (a, c). The strongly reduced induced gap $\Delta_\mathrm{ind}$ is visible in all four conductance matrix elements and closes around $B_{||}\approx\SI{1}{\tesla}$. }
\end{figure}

A measurement of the local and nonlocal conductances a function of magnetic field $B_{||}$ is shown in Fig.~\ref{fig:fieldscan_dev_2}. The gate $\mathrm{W_M}$ was set to \SI{-4.67}{\volt}, while $V_\mathrm{WL}=V_\mathrm{WR}=\SI{-6}{\volt}$. This creates a modulation in the electron density along the elongated NW axis. Both local conductances in Fig.~\ref{fig:fieldscan_dev_2}(a, c) show a parent superconducting gap of \SI{0.25}{\milli\volt} at zero magnetic field which decreases with increasing magnetic field. Several subgap states are visible already at zero magnetic field. The induced gap $\Delta_\mathrm{ind}$ of the proximitized system given by the energy of the lowest ABSs is $\approx \SI{80}{\micro\eV}$ at zero magnetic field. The strongly reduced $\Delta_\mathrm{ind}$ can be interpreted as a result of the lower interface transparency between semiconductor and superconductor of the heterostructure used. With increasing magnetic field, the induced gap $\Delta_{\mathrm{ind}}$ closes at a value of $B_{||}\approx\SI{1}{\tesla}$ due to the ABSs moving to zero energy.\\

In the nonlocal conductances $G_\mathrm{LR}$ and $G_\mathrm{RL}$  [Fig.~\ref{fig:fieldscan_dev_2} (b, d)], the closing of the induced gap is visible at $B_{||}\approx\SI{1}{\tesla}$ where the low energy ABSs merge at zero bias. Higher excited subgap states are resolved in the nonlocal conductance. At voltages above the parent gap $|eV_\mathrm{SD}|>\Delta$ finite nonlocal conductance is visible.\\

\subsection{Measurement of local and nonlocal conductance: gate voltage dependence}

\begin{figure}[h]
\includegraphics[scale=0.85]{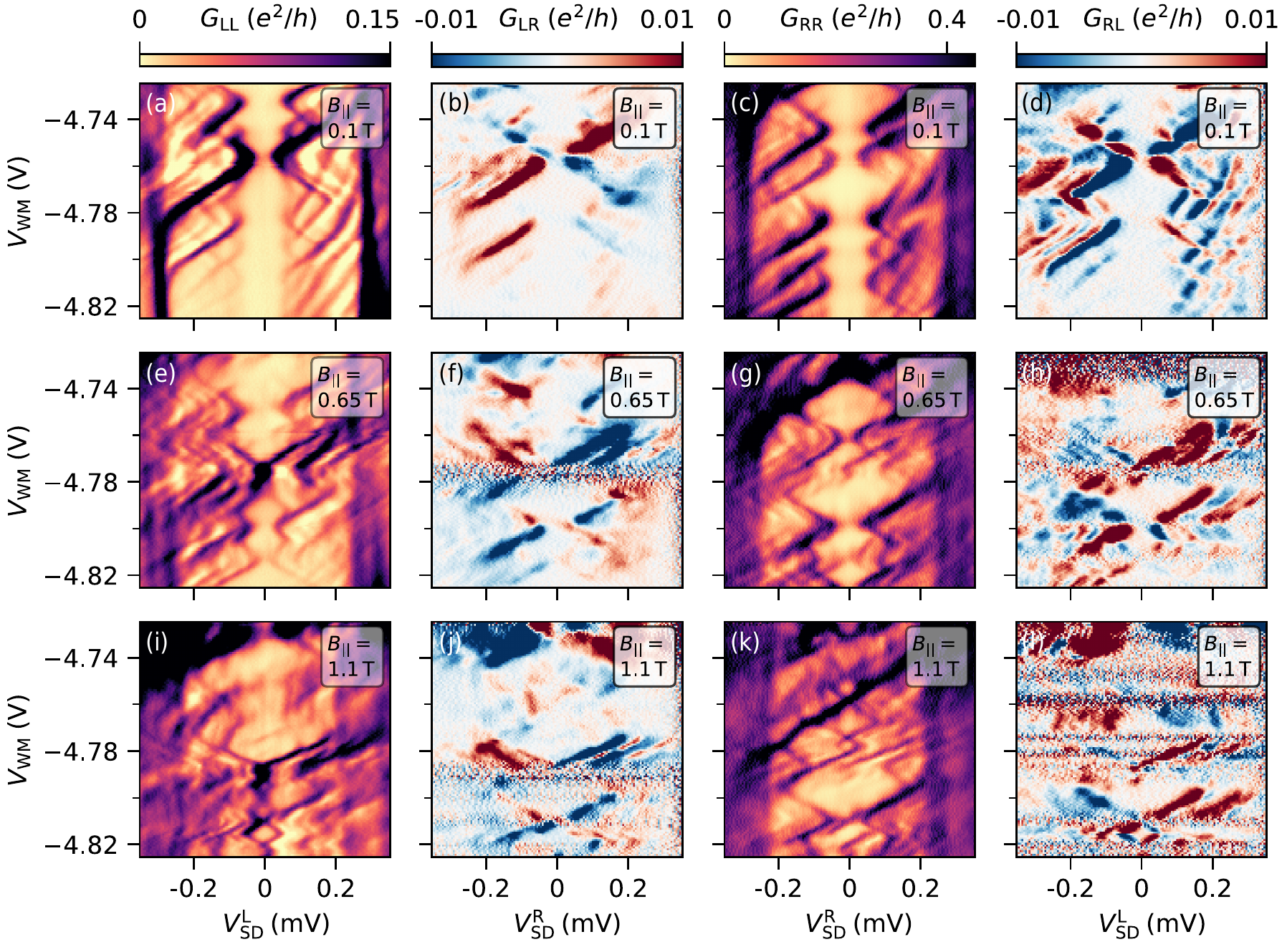}
\caption{\label{fig:plungerscans_1_dev_2}Measured conductance matrix elements at different magnetic field values. (a-d) the induced gap is visible in local and nonlocal conductances. The excited state spectrum consists of discrete states due to the finite size of the NW segment. (i-l) states have crossed zero bias and a low-energy state that oscillates in energy is visible.}
\end{figure}

The different tunneling conductances as a function of gate voltage $V_\mathrm{WM}$ at four distinct magnetic field values are shown in Fig.~\ref{fig:plungerscans_1_dev_2}. At low magnetic field $B_{||}=\SI{0.1}{\tesla}$, there is a finite induced gap $\Delta_\mathrm{ind}$ in the proximitized system. For an infinitely long wire, one expects a continuum of states at energies above the induced gap $\Delta_\mathrm{ind}$. Here, states are confined to a NW segment of length \SI{0.8}{\micro\meter}, which leads to a spectrum of discrete ABSs due to finite size effects instead of a continuum. These discrete states appear clearly as a dense spectrum of intersecting lobes at voltages $\Delta_\mathrm{ind}\leq e V_\mathrm{SD}<\Delta$ in Fig.~\ref{fig:plungerscans_1_dev_2}(a, c). At the voltage $eV_\mathrm{SD}=\Delta_\mathrm{ind}\approx\SI{80}{\micro\eV}$ the states go through a minimum in energy. Similar excited state spectra have been observed in numerical simulations \cite{Mishmash2016, dassarma_goodbadugly}. The presence of disorder, localized bound states in the NW, and multiple transverse modes may further affect the excited state spectrum. The lobe shaped ABSs lead to nonlocal conductance signal as seen from Fig.~\ref{fig:plungerscans_1_dev_2}(b, d) with changes in sign around their minimal energy $\approx\SI{80}{\micro\eV}$.\\

With increasing magnetic field, the ABSs lower their energy. At a magnetic field value $B_{||}=\SI{0.65}{\tesla}$ some of the ABSs merge at zero bias [Fig.~\ref{fig:plungerscans_1_dev_2} (e, g)]. At $B_{||}=\SI{1.1}{\tesla}$ the ABSs have crossed zero bias [see Fig.~\ref{fig:plungerscans_1_dev_2}(i, k)]. Around $V_\mathrm{WM}=\SI{-4.8}{\volt}$, subgap states oscillate around zero bias in Fig.~\ref{fig:plungerscans_1_dev_2}(i, k). Most subgap states in Fig.~\ref{fig:plungerscans_1_dev_2} appear both in the local and nonlocal conductances, which is a strong indication that they stem from extended bound states that couple to the normal leads at both ends of the NW segment.\\

\begin{figure}[h!]
\includegraphics[scale=0.85]{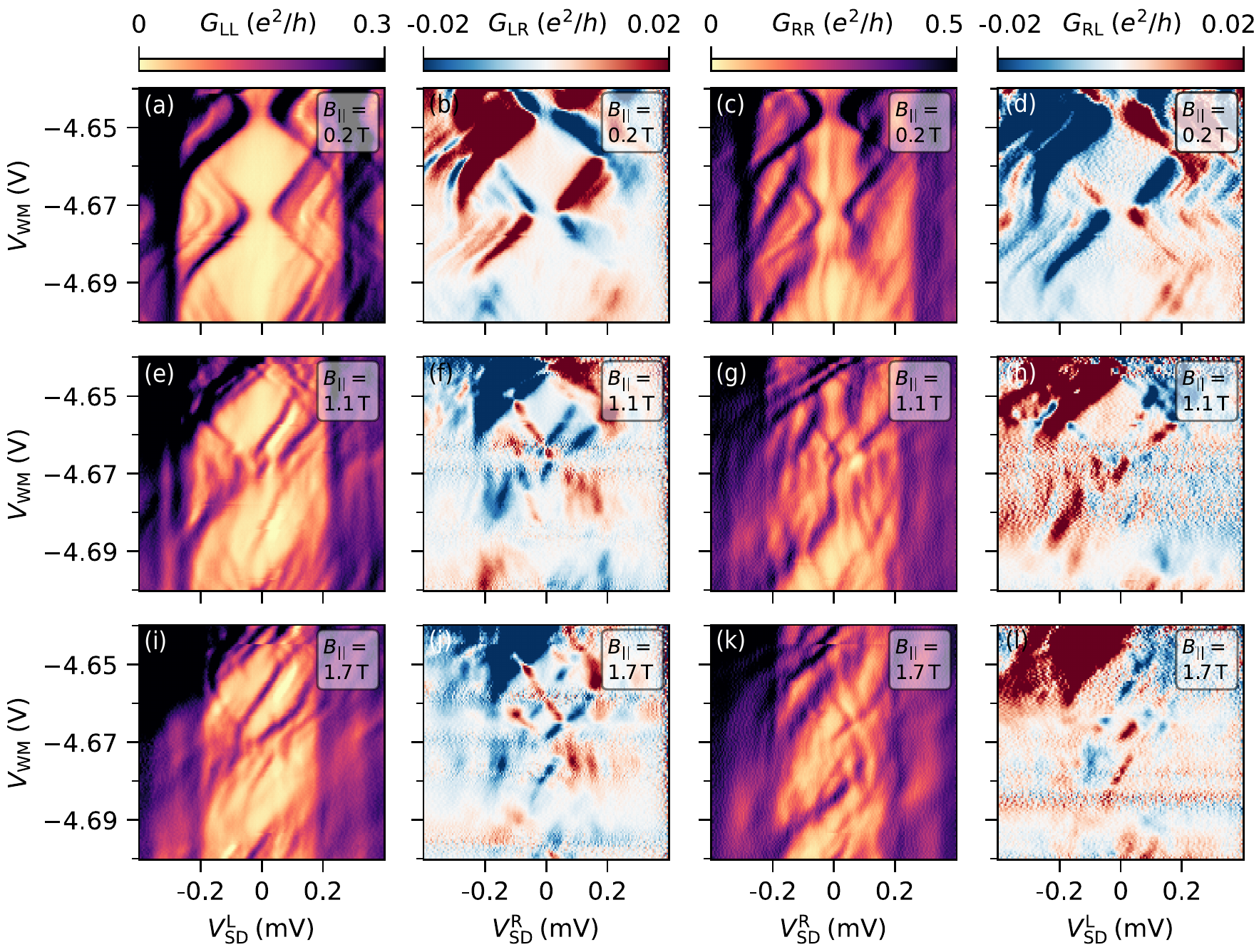}
\caption{\label{fig:plungerscans_2_dev_2}Local and nonlocal conductances at different values of $B_{||}$. The subgap state close to zero bias in (c) is not visible in the three other conductances measured at the same field value. Some of the lobe shaped ABSs appear in both local conductances (a, c) and in the nonlocal conductances (b, d) with the characteristic sign changes. (e-h) subgap state around $V_\mathrm{WM}=\SI{-4.67}{\volt}$ oscillates around zero bias and is visible in all local and nonlocal conductances.}
\end{figure}

Local and nonlocal conductances at less negative gate voltage $V_\mathrm{WM}$ are shown in Fig.~\ref{fig:plungerscans_2_dev_2}. At low magnetic field $B_{||}=\SI{0.2}{\tesla}$ subgap states due to finite size confinement appear in the local and nonlocal conductances [Fig.~\ref{fig:plungerscans_2_dev_2}(a-d)]. The nonlocal conductance signal of individual subgap states furthermore undergoes the characteristic sign changes at points where the state reaches a minimum in energy ($\approx\SI{80}{\micro\eV}$) or where subgap states cross ($\approx \SI{0.2}{\milli\eV}$). The measurement of $G_\mathrm{RR}$ furthermore shows a subgap state around zero bias. This state does not appear in the measurement of $G_\mathrm{LL}$ suggesting that it does not couple to both leads at this magnetic field value. The state consequently is not visible in the nonlocal conductances [Fig.~\ref{fig:plungerscans_2_dev_2}(b, d)]. At a higher magnetic field value of $B_{||}=\SI{1.1}{\tesla}$ two subgap states merge forming a zero bias peak in the measurement of $G_\mathrm{RR}$ [Fig.~\ref{fig:plungerscans_2_dev_2}(b)]. This state appears only for a small range of gate voltage around $V_\mathrm{WM}=\SI{-4.67}{\volt}$ in $G_\mathrm{LL}$ in Fig.~\ref{fig:plungerscans_2_dev_2}. In this range, the two states cross zero bias and overshoot. This leads to a characteristic signature in nonlocal conductance with a sign change at the point of local maximum in energy of the bound states.
At an even higher magnetic field value $B_{||}=\SI{1.7}{\tesla}$, the conductance acquires a strong asymmetry, making it difficult to track the evolution of individual states in the local conductances.

\subsection{Comparison between antisymmetric parts of local and nonlocal conductances}
\begin{figure}[h!]
\includegraphics[scale=0.85]{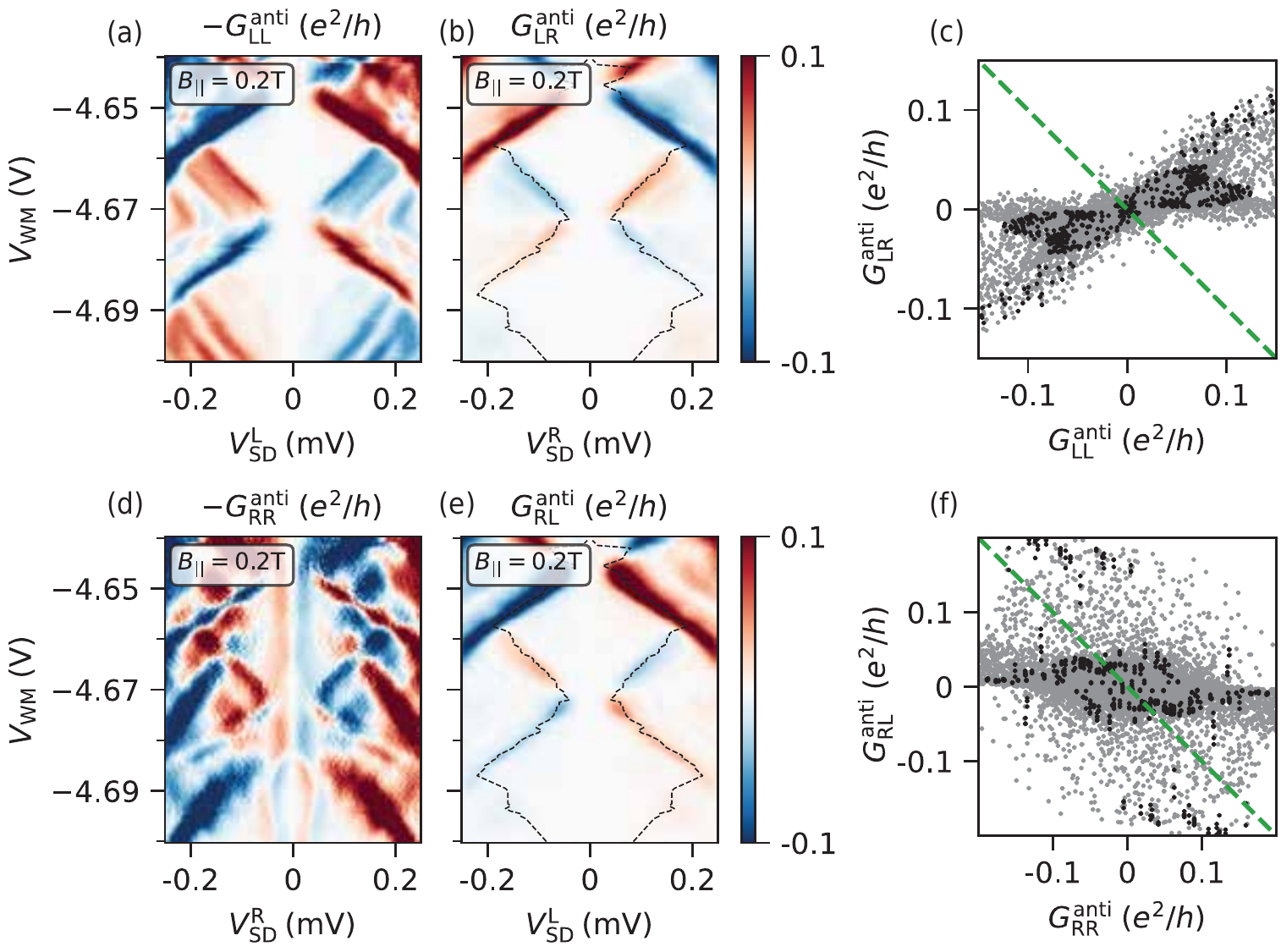}
\caption{\label{fig:antisymmetries_lowB_dev_2}(a, b) Comparison of $-G_\mathrm{LL}^\mathrm{anti}$ and $G_\mathrm{LR}^\mathrm{anti}$. (c) parametric plot of the data in (a, b). The black data points originate from a three pixel wide window around the black dashed line in (b). (d, e) shows a comparison of $-G_\mathrm{RR}^\mathrm{anti}$ and $G_\mathrm{RL}^\mathrm{anti}$. (f) Parametric plot of the data in (d, e). The black data points originate from a three pixel wide window around the black dashed line in (e).}
\end{figure}

Similar to the the analysis in section \ref{sec:antisymmetries_dev1}, the antisymmetric parts of the conductance matrix elements are calculated from the data in Fig.~\ref{fig:plungerscans_2_dev_2}. The quantities $-G_\mathrm{LL}^\mathrm{anti}$ and $G_\mathrm{LR}^\mathrm{anti}$ at $B_{||}=\SI{0.2}{\tesla}$ are shown in Fig.~\ref{fig:antisymmetries_lowB_dev_2}(a, b). While the two quantities are expected to be equal according to Eq.~\ref{eq:anti-symmetry} predicted by theory \cite{karsten_nl_spectroscopy}, the experimental data shows that $-G_\mathrm{LL}^\mathrm{anti}$ has a larger magnitude and different sign than $G_\mathrm{LR}^\mathrm{anti}$. This lack of similarity between the data is also visible in the parametric plot in Fig.~\ref{fig:antisymmetries_lowB_dev_2}(c), where all data points from panels (a) and (b) are plotted with the data points taken in a three pixel window around the state shown as a dashed black line in (b).

Between $-G_\mathrm{RR}^\mathrm{anti}$ and $G_\mathrm{RL}^\mathrm{anti}$ there is a large discrepancy visible in Fig.~\ref{fig:antisymmetries_lowB_dev_2}(d, e). A particularly large antisymmetric component of the local conductance $G_\mathrm{RR}$ is visible at low values of source-drain bias $|V_\mathrm{SD}^\mathrm{R}|\leq\SI{80}{\micro\V}$ due to the presence of the low energy subgap state that only appears in $G_\mathrm{RR}$ as discussed earlier. A parametric plot of the data in (d, e) is shown in (f) and confirms the lack of points close to the expected theory relation given by the green dashed line.

\begin{figure}[h!]
\includegraphics[scale=0.85]{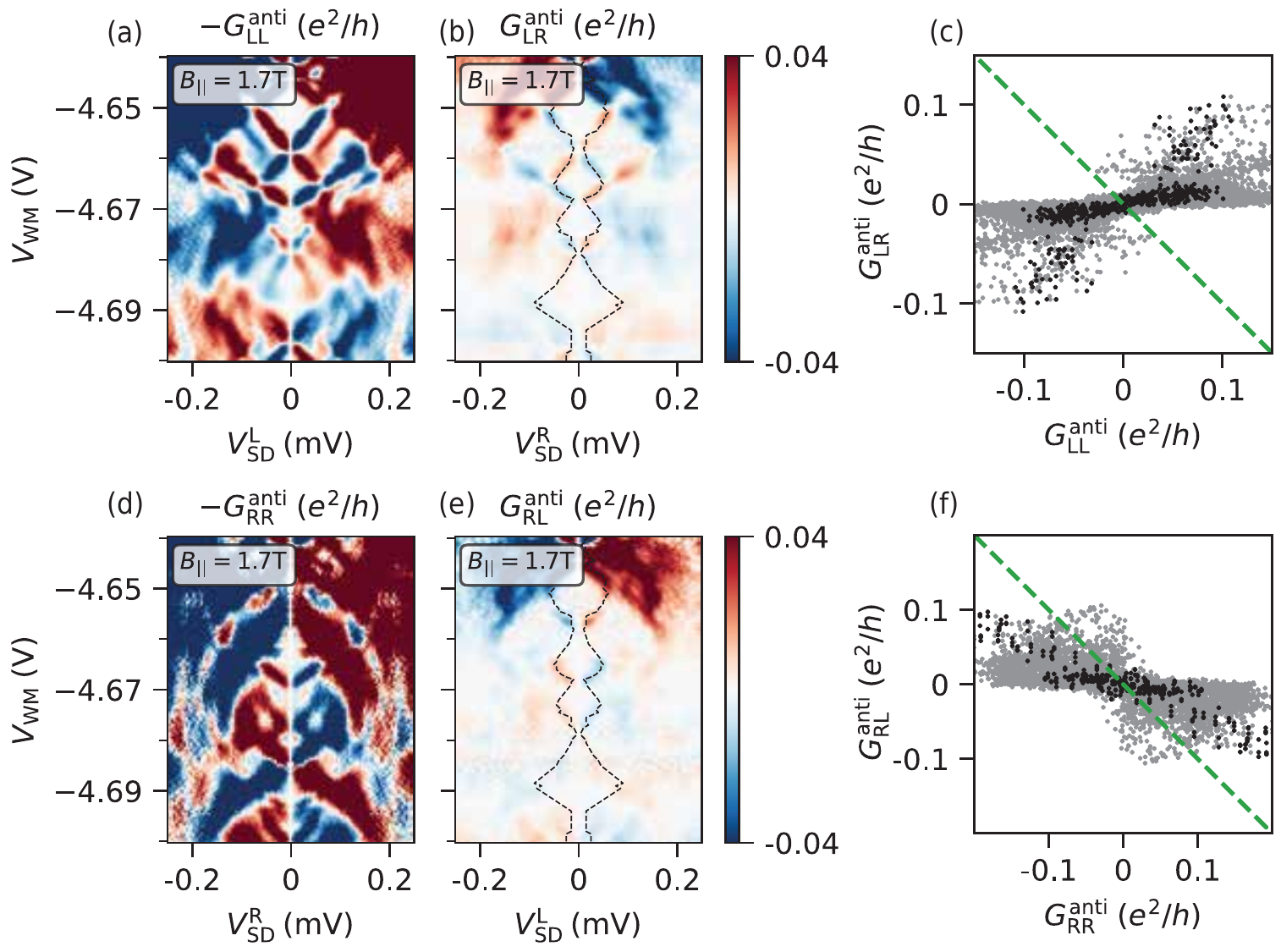}
\caption{\label{fig:antisymmetries_highB_dev_2}(a, b) Comparison of $-G_\mathrm{LL}^\mathrm{anti}$ and $G_\mathrm{LR}^\mathrm{anti}$. (c) Parametric plot of the data in (a,b). The green dashed line indicates the relation between data points expected from theory and the black data points originate from three pixel wide window around the black dashed line in (b). (d, e) comparison of $-G_\mathrm{RR}^\mathrm{anti}$ and $G_\mathrm{RL}^\mathrm{anti}$. (f) Parametric plot of the data in (d, e).  The green dashed line indicates the relation between data points expected from theory and the black data points originate from three pixel wide window around the black dashed line in (e).}
\end{figure}

In Fig.~\ref{fig:antisymmetries_highB_dev_2} the antisymmetric component of the conductances measured at $B_{||}=\SI{1.7}{\tesla}$ is shown. There is no equality of $-G_\mathrm{LL}^\mathrm{anti}$ and $G_\mathrm{LR}^\mathrm{anti}$. The same holds true for $-G_\mathrm{RR}^\mathrm{anti}$ and $G_\mathrm{RL}^\mathrm{anti}$.

\begin{figure}[h!]
\includegraphics[scale=0.85]{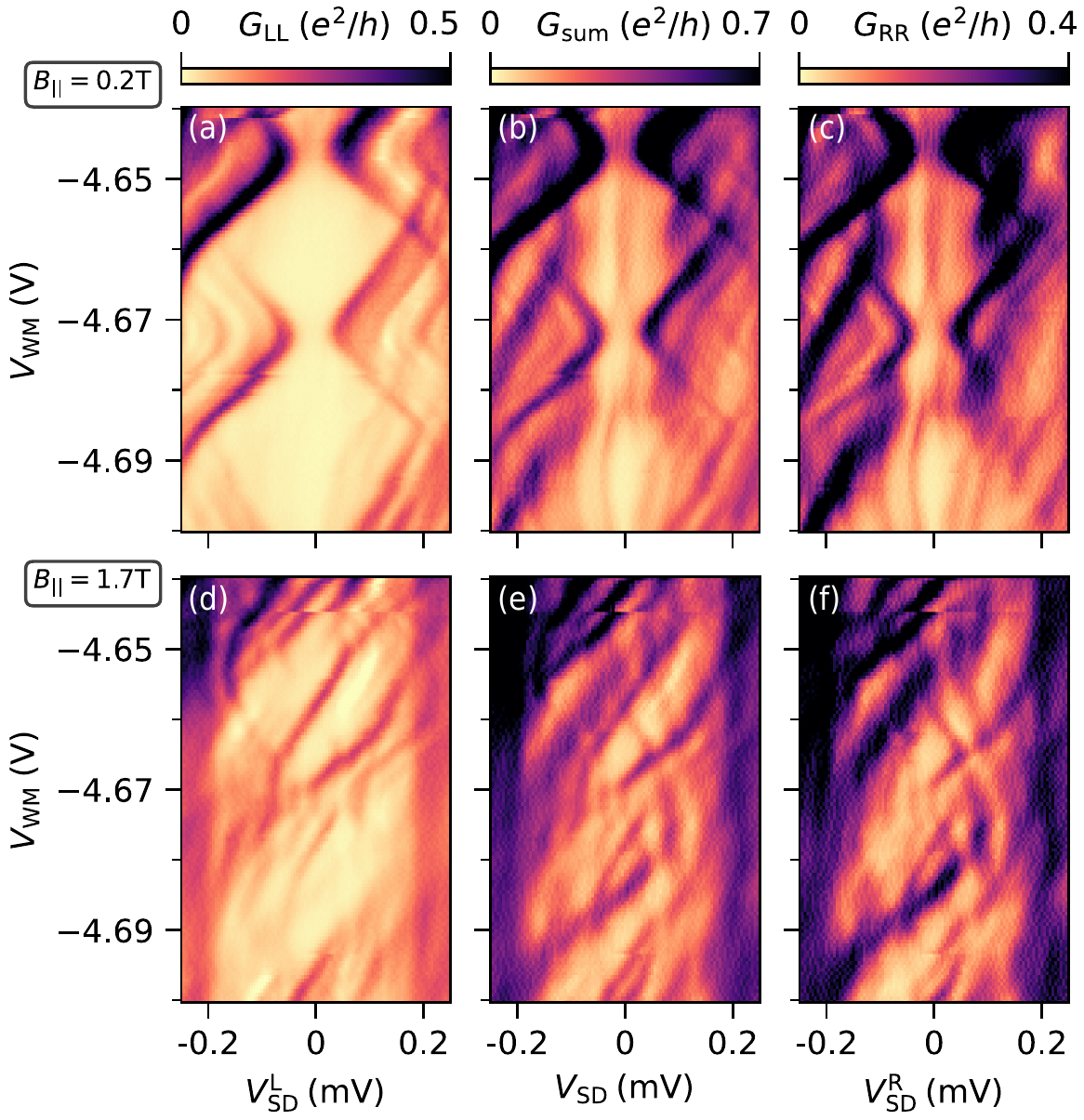}
\caption{\label{fig:symmetries_lowB_dev_2}(a) Tunneling conductance $G_\mathrm{LL}$, (b) the sum of all conductance matrix elements $G_\mathrm{sum}$, (c) tunneling conductance $G_\mathrm{RR}$ measured at $B_{||}=\SI{0.2}{\tesla}$. A notable asymmetry of $G_\mathrm{sum}$ with respect to $V_\mathrm{SD}$ is visible. (d) tunneling conductance $G_\mathrm{LL}$, (e) the sum of all conductance matrix elements $G_\mathrm{sum}$, (f) tunneling conductance $G_\mathrm{RR}$ measured at $B_{||}=\SI{1.7}{\tesla}$.}
\end{figure}

As a consequence of the experimental data not fulfilling Eq.~\ref{eq:anti-symmetry}, forming the sum $G_\mathrm{sum}$ of all conductance matrix elements does not recover a symmetric function with respect to $V_\mathrm{SD}$ compared to the local conductances. This can be clearly seen for $B_{||}=\SI{0.2}{\tesla}$ in Fig.~\ref{fig:symmetries_lowB_dev_2}(a-c) and for $B_{||}=\SI{1.7}{\tesla}$ in Fig.~\ref{fig:symmetries_lowB_dev_2}(d-f) where $G_\mathrm{LL}$, $G_\mathrm{sum}$, and $G_\mathrm{RR}$ are plotted side-by-side.\\

While we found that the lowest energy subgap state measured on device 1 fulfills the relation given by Eq.~\ref{eq:anti-symmetry}, the data for device 2 do not follow this relation. Several reasons for Eq.~\ref{eq:anti-symmetry} being violated have been given in previous works \cite{gerbold_nonlocal,melo_asym}. The two devices presented here also differ in the composition of their underlying hybrid heterostructure and in the NW length.

\subsection{Extracted values of $Q_\mathrm{L}$ and $Q_\mathrm{R}$}

\begin{figure}[h!]
\includegraphics[scale=0.85]{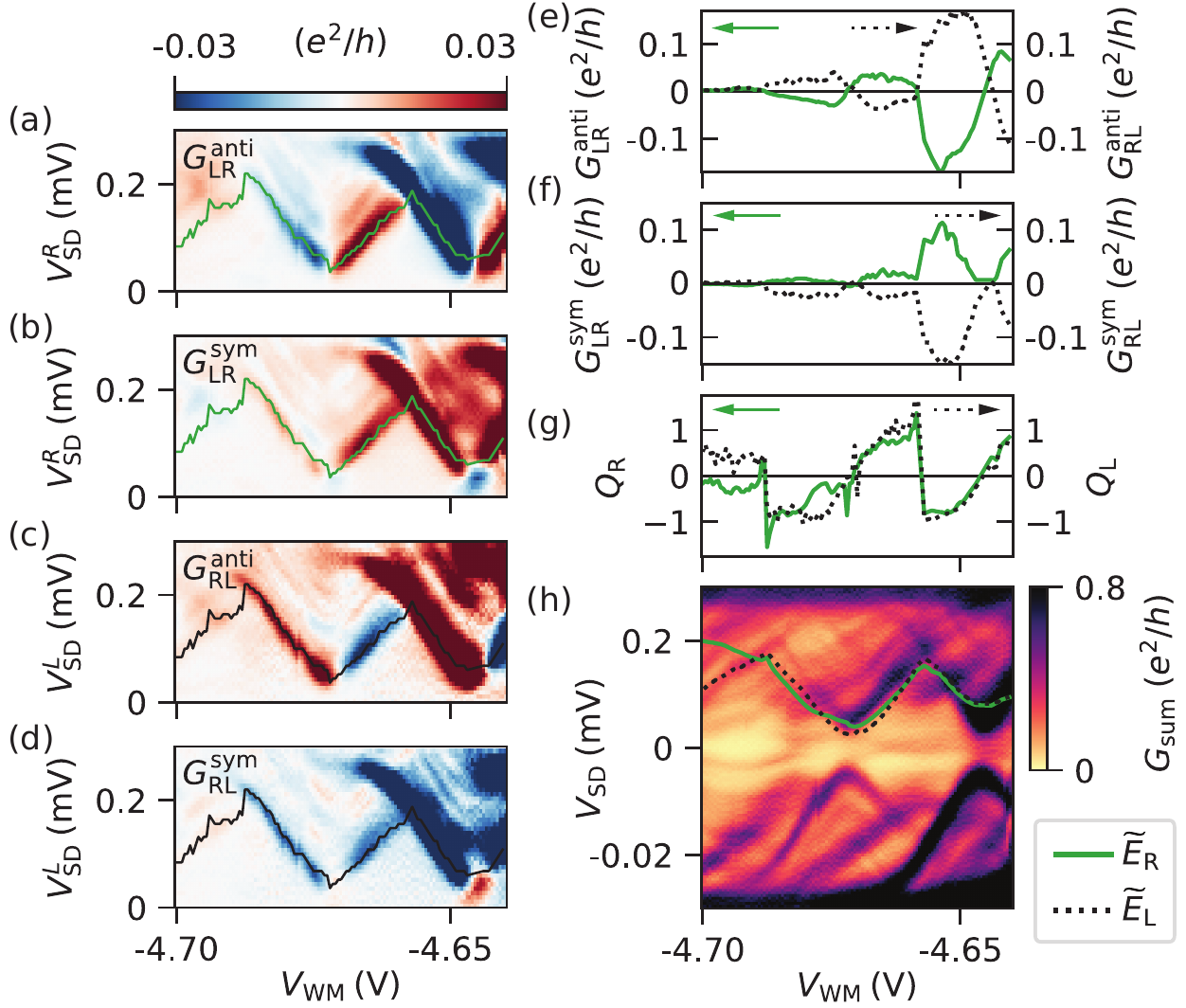}
\caption{\label{fig:Q_extr_dev_2_0.2T}(a-d) Symmetric and antisymmetric components of the nonlocal conductances measured at $B_{||}=\SI{0.2}{\tesla}$. The values at the position of the green and black lines are plotted in (e, f) and were used to extract $Q_\mathrm{L}$, $Q_\mathrm{R}$, which are shown in (g). The two quantities show a continuous transition from -1 to +1 where the ABS reaches a minimum in energy. Sharp transitions are visible at the crossing of two states at $V_\mathrm{SD}=\SI{0.2}{\milli\volt}$. (h) shows the sum of all conductances $G_\mathrm{sum}$. The dotted black, solid green display the energies $\widetilde{E}_\mathrm{L}$, $\widetilde{E}_\mathrm{R}$ inferred from the integrated $Q_\mathrm{L}$, $Q_\mathrm{R}$.}
\end{figure}

Figure \ref{fig:Q_extr_dev_2_0.2T}(a-d) shows the symmetric and antisymmetric component of the nonlocal conductances $G_\mathrm{LR}$ and $G_\mathrm{RL}$ in Fig.~\ref{fig:plungerscans_2_dev_2}(b, d) measured at $B_{||}=\SI{0.2}{\tesla}$. The value for the lowest excited state is shown in \ref{fig:Q_extr_dev_2_0.2T}(e, f). The resulting values of $Q_\mathrm{L}$ and $Q_\mathrm{R}$ are shown in Fig.~\ref{fig:Q_extr_dev_2_0.2T}(g). The curves for $Q_\mathrm{L}$ and $Q_\mathrm{R}$ are similar. At the value where the ABS goes through a minimum in energy, $Q_\mathrm{j}$ ($j\in\{\mathrm{L},\,\mathrm{R}\}$) undergoes a continuous crossover from negative to positive values. At points where the ABS intersects with a neighboring ABS around $V_\mathrm{SD}\approx\SI{0.2}{\milli\volt}$ an abrupt change from $Q_\mathrm{j}>0$ to $Q_\mathrm{j}<0$ is visible. Regions of positive $Q_\mathrm{j}$ coincide with regions where the state energy has a positive slope with respect to the gate voltage $V_\mathrm{WM}$. Vice versa, regions of negative $Q_\mathrm{j}$ coincide with regions where the state energy has a negative slope. This is in agreement with the interpretation of $Q_\mathrm{j}$ reflecting the charge character of the bound state. 

Integrating $Q_\mathrm{L}$ and  $Q_\mathrm{R}$, taking into account a linear background and rescaling by a constant lever arm according to Eq.~\ref{eq:Einf} leads to the curves shown in Fig.~\ref{fig:Q_extr_dev_2_0.2T}(h). The curve matches the state energy over a large range of gate voltage $V_\mathrm{WM}$, which suggests that $Q_\mathrm{L}$, $Q_\mathrm{R}$ are proportional to the total, integrated charge of the ABS \cite{karsten_nl_spectroscopy}. 

\begin{figure}[h!]
\includegraphics[scale=0.85]{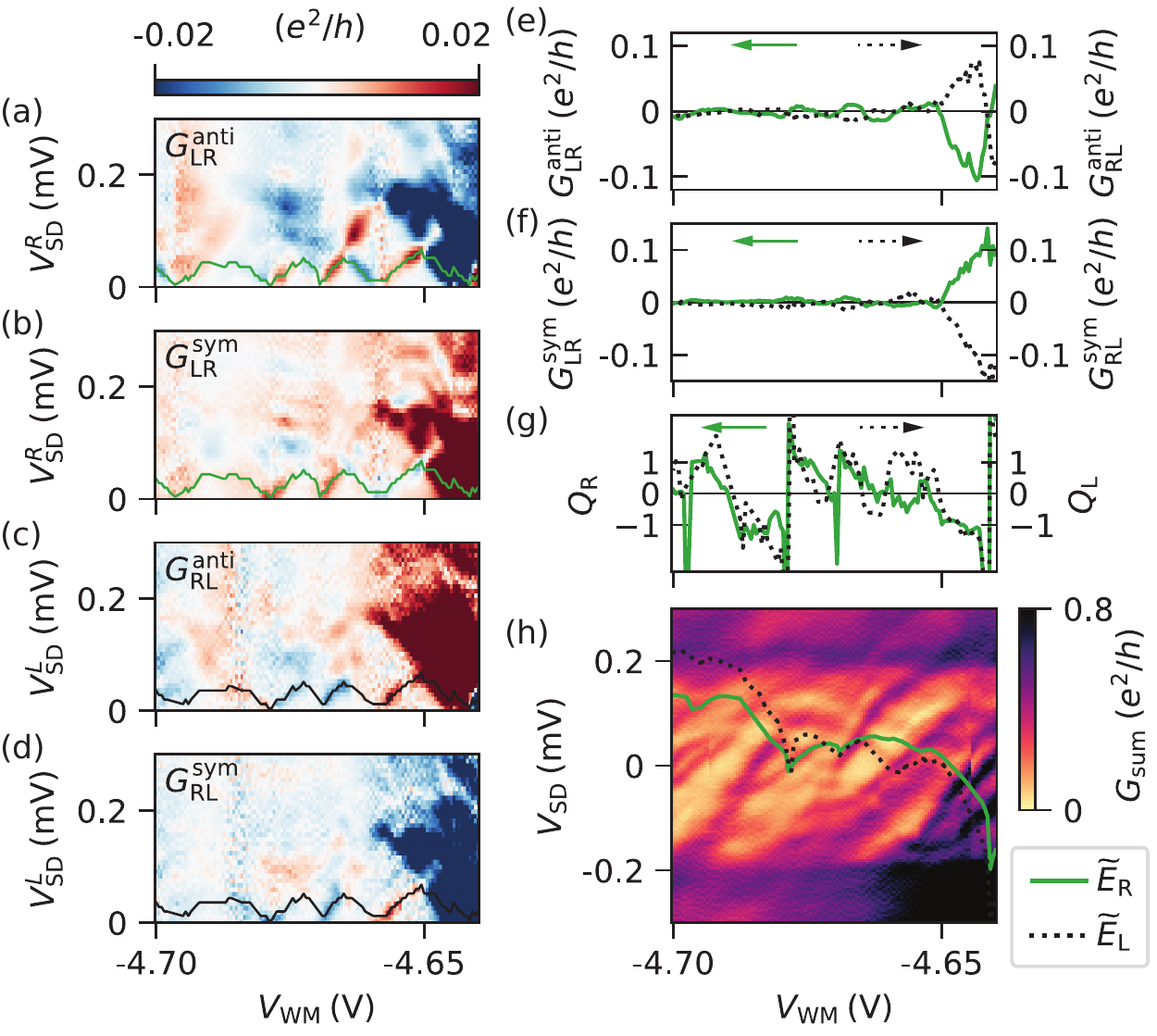}
\caption{\label{fig:Q_extr_dev_2_1.7T}(a-d) Symmetric and antisymmetric components of the nonlocal conductances measured at $B_{||}=\SI{1.7}{\tesla}$. The values at the position of the green and black lines are plotted in (e, f) and were used to extract $Q_\mathrm{L}$, $Q_\mathrm{R}$, which are plotted in (g). (h) shows the sum of all conductances $G_\mathrm{sum}$. The dotted black, solid green lines in represent $\widetilde{E}_\mathrm{L}$, $\widetilde{E}_\mathrm{R}$ inferred from the integrated $Q_\mathrm{L}$, $Q_\mathrm{R}$.}
\end{figure}

For the nonlocal conductance data in Fig.~\ref{fig:plungerscans_2_dev_2}(i-l) at $B_{||}=\SI{1.7}{\tesla}$, $Q_\mathrm{L}$ and $Q_\mathrm{R}$ were extracted. The underlying symmetric and antisymmetric components are plotted in Fig.~\ref{fig:Q_extr_dev_2_1.7T}(a-f). The resulting $Q_\mathrm{L}$ and $Q_\mathrm{R}$ are plotted in Fig.~\ref{fig:Q_extr_dev_2_1.7T}(g). The two curves show a similar evolution with $V_\mathrm{WM}$. The inferred energy $\widetilde{E}_\mathrm{L}$, $\widetilde{E}_\mathrm{R}$ in Fig.~\ref{fig:Q_extr_dev_2_1.7T}(h) does not match the energy of the low energy subgap state. The extracted values $Q_\mathrm{j}$ are the quotient of two experimentally measured quantities subjected to noise. The antisymmetric part enters this quotient as denominator and is zero at $V_\mathrm{SD}=\SI{0}{\volt}$ by construction. Consequently, the extraction of $Q_\mathrm{j}$ is very sensitive for states close or at $V_\mathrm{SD}=\SI{0}{\volt}$. This and the overall small signal may lead to a deviation between the energy of the state and $\widetilde{E}_\mathrm{L}$, $\widetilde{E}_\mathrm{R}$ in the case shown in Fig.~\ref{fig:Q_extr_dev_2_1.7T}(h).

\end{document}